\documentclass[11pt,draftclsnofoot,peerreviewca,letterpaper,onecolumn]{IEEEtran}

\usepackage{cite}
\usepackage{graphicx,color,epsfig,rotating}
\usepackage{amsfonts,amsmath,amssymb,bbm}
\usepackage{algorithm, algorithmic}
\usepackage{subfigure}
\usepackage{amsmath}
\usepackage{cite}
\usepackage{mdwtab}
\usepackage{subfigure}
\usepackage{placeins}
\usepackage{psfrag, graphicx}
\usepackage[latin1]{inputenc}
\usepackage{amssymb}
\usepackage{makeidx}

\long\def\comment#1{}

\newfont{\bbb}{msbm10 scaled 700}

\newfont{\bb}{msbm10 scaled 1100}

\newcommand{\PP}{\mbox{\bb P}}

\newcommand{\FF}{\mbox{\bb F}}

\newcommand{\EE}{\mbox{\bb E}}

\newcommand{\Ac}{{\cal A}}

\newcommand{\Dc}{{\cal D}}
\newcommand{\Ec}{{\cal E}}
\newcommand{\Fc}{{\cal F}}
\newcommand{\Gc}{{\cal G}}

\newcommand{\Kc}{{\cal K}}
\newcommand{\Lc}{{\cal L}}
\newcommand{\Mc}{{\cal M}}

\newcommand{\Sc}{{\cal S}}

\newcommand{\Uc}{{\cal U}}

\newcommand{\Vc}{{\cal V}}

\newcommand{\Zc}{{\cal Z}}

\newcommand{\fsf}{{\sf f}}

\newcommand{\tsf}{{\sf t}}

\newcommand{\Asf}{{\sf A}}

\newcommand{\eqdef}{\stackrel{\Delta}{=}}

\newcommand{\be}{\begin{equation}}
\newcommand{\ee}{\end{equation}}
\newcommand{\bea}{\begin{eqnarray}}
\newcommand{\eea}{\end{eqnarray}}

\def\fsf{ {\sf f}}

\newtheorem{defn}{Definition}%[section]
\newtheorem{theorem}{Theorem}%[section]
\newtheorem{corollary}{Corollary}%[section]
\newtheorem{remark}{Remark}%[section]

\begin{document}

\setcounter{page}{1}

\title{Towards Finite File Packetizations in Wireless Device-to-Device Caching Networks}

\author{Nicholas Woolsey,~\IEEEmembership{Student Member,~IEEE}, Rong-Rong Chen,~\IEEEmembership{Member,~IEEE},\\ and Mingyue Ji,~\IEEEmembership{Member,~IEEE}
\thanks{The authors are with the Department of Electrical Engineering,
University of Utah, Salt Lake City, UT 84112, USA. (e-mail: nicholas.woolsey@utah.edu, rchen@ece.utah.edu and mingyue.ji@utah.edu)}
}

\maketitle

\vspace{-0.5cm}

\begin{abstract}
We consider wireless device-to-device (D2D) caching networks with single-hop transmissions. Previous work has demonstrated that caching and coded multicasting can significantly increase per user throughput. However, the state-of-the-art coded caching schemes for D2D networks are generally impractical because content files are partitioned into an exponential number of packets with respect to the number of users if both library and memory sizes are fixed. In this paper, we present two combinatorial approaches of D2D coded caching network design with reduced packetizations and desired throughput gain compared to the conventional uncoded unicasting. The first approach uses a ``hypercube'' design, where each user caches a ``hyperplane" in this hypercube and the intersections of ``hyperplanes" represent coded multicasting codewords. In addition, we extend the hypercube approach to a decentralized design. The second approach uses the Ruzsa-Szem\'eredi graph to define the cache placement. Disjoint matchings on this graph represent coded multicasting codewords. Both approaches yield an exponential reduction of packetizations while providing a per-user throughput that is comparable to the state-of-the-art designs in the literature. Furthermore, we apply spatial reuse to the new D2D network designs to further reduce the required packetizations and significantly improve per user throughput for some parameter regimes.

\end{abstract}


%\newpage

\section{Introduction}
\label{section: intro}
Wireless caching is a promising approach to significantly improve the user throughput %reduce traffic load
and simultaneously accommodate a large number of %simultaneous
user demands in future generations of wireless networks \cite{maddah2014fundamental,bastug2014living,paschos2016caching,liu2016caching,shanmugam2014finite,wan2016caching,wan2016optimality,yu2017characterizing,Karamchandani16,tandon2017improved,Jeon2017wireless,liu2016cache,shariatpanahi2017physical}. %wan2017combination
In this paper, we investigate achievable {\em coded caching} schemes in device-to-device (D2D) caching networks, %Similar to the shared-link schemes,
where users strategically cache packets of content files to enable coded multicasting which serves distinct content to multiple users with one channel use. Different from the seminal {\em shared link} caching networks \cite{maddah2014fundamental}, where one source node (base station) with access to the entire library serves all the users over a multicast channel, in D2D networks, users receive requested packets from other users. For such a network consisting of $n$ users, each caching an equivalent $M$ files out of a library of $m$ files, previous work demonstrates that if $Mn\geq m$ and $m\geq n$ and when spatial reuse is not allowed, meaning that any transmission can be successfully received by any users in the network, the transmission rate (i.e., normalized traffic load)  %D2D caching networks
is $\Theta\left(\frac{m}{M}\right)$, which is not a function of $n$. Hence,  the aggregate throughput of the network is scalable \cite{ji2016fundamental}.\footnote{Note that when no spatial reuse is allowed, the per user throughput is inversely proportional to the traffic load in the network.} This surprising result shows that the transmission rate of the {shared link} caching scheme in \cite{maddah2014fundamental} and D2D caching scheme in \cite{ji2016fundamental} are identical for a large number of users.\footnote{We will use the following standard ``order'' notation: given two functions $f$ and $g$, we say that: 1)  $f(n) = O\left(g(n)\right)$ if there exists a constant $c$ and integer $N$ such that  $f(n)\leq cg(n)$ for $n>N$. 2) $f(n)=o\left(g(n)\right)$ if $\lim_{n \rightarrow \infty}\frac{f(n)}{g(n)} = 0$.
3) $f(n) = \Omega\left(g(n)\right)$ if $g(n) = O\left(f(n)\right)$. 4)
$f(n) = \omega\left(g(n)\right)$ if $g(n) = o\left(f(n)\right)$.
5) $f(n) = \Theta\left(g(n)\right)$ if $f(n) = O\left(g(n)\right)$ and~$g(n) = O\left(f(n)\right)$.}

D2D caching networks have the potential to provide some unique advantages. %The most obvious advantage is that
For example, D2D caching networks have a greater flexibility to implement spatial reuse in comparison to shared link caching networks. %is independent of the location of base stations.
The authors in \cite{ji2016fundamental} demonstrate that users in a D2D network can be grouped into clusters based on proximity. %to form many smaller caching networks. %. Based on the protocol model \cite{xue2006scaling},
%Multiple clusters can perform the coded multicasting delivery simultaneously and without interference, taking advantage of spatial reuse.  %Assuming not all clusters can be actively transmitting simultaneously,
%While clustering increases the transmission rate for the overall network% increases the transmission rate,
%, clustering is to shown increase
%Hence, the per-user throughput can be improved since the link rate in each cluster %since the link quality is relatively better for users of spatially smaller networks. Furthermore, the caching phase of D2D networks can accommodate a {\em shared link} delivery phase with coded multicasts. This means by using a D2D caching phase, the delivery phase of can be performed by a {\em shared link}, D2D communication, or a combination of the two, providing more options to optimize the spectral efficiency of the overall network.
These clusters can perform the coded multicasting delivery simultaneously and surprisingly,  %Assuming not all clusters can be actively transmitting simultaneously,
the order-optimal traffic load in each cluster is identical to the traffic load when no clustering is enabled (e.g., consider the entire network as a single cluster). % increases the transmission rate,
Nevertheless, clustering may improve per user throughput since the link rate (bits/second/Hz) in each cluster may increase as the size of each cluster decreases. Due to their unique characteristics, the study of the fundamental limits of D2D caching networks has become a popular topic in the past few years  \cite{ji2015tradeoff,Ji2016WirelessD2D,ji2016fundamental,tandon2017improved,ji2017multihop,wan2018novel,zewail2018device,yapar2019optimality,ibrahim2019device,lee2019d2d,wan2019device}.

The promised gain in per user throughput of the state-of-the-art coded D2D caching schemes relies on a large amount of file packetization which makes the networks impractical to implement. Files need to be split into a very large number of packets and therefore the files will be unrealistically large for many caching network implementations. Moreover, %a recent study which implements a shared link caching network with hardware demonstrates that
to implement a coded caching network, a header packet used for labeling the content of the multicast, must be included with every transmission \cite{fadlallah2017coding}. The size of this header will become large as the number of packets per file increases which could significantly increase the rate. In this paper, we study and propose new achievable coded caching schemes in D2D networks %cache placement, coded multicasting and scheduling schemes
such that the packetization of each file is significantly reduced without sacrificing much throughput of the currently proposed D2D caching schemes.%Furthermore, only so many partitions are possible for files with a finite number of bits. For these reasons, the scheme in \cite{ji2016fundamental} is impractical to implement for a large number of users

%Spatial reuse in D2D caching networks partially solves this issue by forming many small user clusters where the number of file partitions decreases as the number of clusters increases \cite{ji2016fundamental}. However, spatial reuse alone may not be able to reduce the packetization sufficiently since the number of clusters in the network can be limited.

%One of the limitations of the achievable D2D scheme in \cite{ji2016fundamental} is the requirement of the number of packets per file when there is no (or limited) spatial reuse.
%In particular, if no spatial reuse is considered, let $t=\frac{nM}{m} \in \mathbb{Z}^+$, the required number of packets per file, $K=t {n \choose t}$, which grows exponentially as the number of users, $n$,  increases.
%A recent study which implemented a caching network with hardware demonstrated that a header packet, labeling the content of the multicast, must be included with every transmission \cite{fadlallah2017coding}. The size of this header will become large as the number of packets per file increases. Furthermore, only so many partitions are possible for files with a finite number of bits. For these reasons, the scheme in \cite{ji2016fundamental} is impractical to implement for a large number of users.

\subsection{Related Work and Contributions}

There have been multiple results studying the large packetization issue in shared link caching networks \cite{yan2017placement,tang2018coded,krishnan2019coded} and some results are discussed here.
The work of \cite{yan2017placement} used a Placement Delivery Array (PDA) to investigate new placement and delivery schemes. One of the proposed schemes in \cite{yan2017placement} %designs reduced
reduces the number of users served in each coded multicast transmission by $1$ while significantly reducing the packetization compared to the seminal work of \cite{maddah2014fundamental}. In this way, the rate only increases slightly while greatly increasing the number of practical parameters regimes. Furthermore, the authors of \cite{tang2018coded} demonstrated the connection between $(m,k)$ linear block codes over ${\rm GF}(q)$ with coded caching network design. The generated codeword matrix defines the cache of $mk$ users and at most $(k+1)q^k$ file packetizations are necessary. While the scheme only works for linear block codes with the $(k,k+1)$-consecutive column property (see \cite{tang2018coded}), the authors demonstrated the flexibility of this approach and in some cases it can be used to design a caching network given $n$, $m$ and $M$ while meeting a specific packetization requirement. While the schemes of \cite{yan2017placement} and \cite{tang2018coded} significantly reduce the file packetization compared to \cite{maddah2014fundamental}, all of these schemes require an exponential number of packets per file compared to the number of users. A recent result has demonstrated that caching schemes where a linear number of packets per file are necessary by using a Ruzsa-Szem\'eredi graph \cite{ruzsa1978triple, alon2012nearly} to design %the caching network
coded caching scheme in order to have the global caching gain \cite{maddah2014fundamental,shanmugam2017coded}. While this approach requires a %an arbitrarily
large number of users, it has proven the existence of sub-exponential schemes which inspires the search for practical caching schemes with reduced packetizations.

There has been limited work studying file packetization in D2D caching networks. In %the %seminal
%D2D coded caching work of
\cite{ji2016fundamental}, the authors demonstrate that if no spatial reuse is allowed, let $t=\frac{nM}{m} \in \mathbb{Z}^+$, the required number of packets per file is $K=t {n \choose t}$ which grows exponentially as the number of users, $n$, increases.
%Ultimately, the proposed placement phase of \cite{ji2016fundamental} is essentially the placement phase of the shared link scheme in \cite{maddah2014fundamental}, but packets are simply split into $t$ smaller packets.
The authors of \cite{ji2016fundamental} also explored the concept of user clustering in order to exploit spatial reuse to increase per user throughput. Moreover, %the authors
it was found that clustering also has the potential to reduce packetizations. Another approach to study D2D coded caching networks is the use of a D2D Placement Delivery Array (DPDA) \cite{wang2017placement}. By using the DPDA, the authors first derived a lower bound for the rate of a coded caching network as $R=\frac{m}{M}-1$ and also a lower bound for packetizations when the rate lower bound is met and $t\in\{1,2,n-2,n-1\}$. Furthermore, the work of \cite{wang2017placement} demonstrated that the scheme of \cite{ji2016fundamental} meets the lower bound on rate always, meets the lower bound on packetization for $t\in\{1,K-1\}$ but does not meet the lower bound on packetization when $t\in\{2,K-2\}$. The authors of \cite{wang2017placement} developed a specific scheme for $t=2$ and $t=K-2$ which meets the lower bound on rate and packetization. An open question remains as to the existence of D2D coded caching networks which work for a large range of $t$ and are designed specifically for D2D and not simply adapted from a shared link scheme. Furthermore, only the scheme of \cite{ji2016fundamental} has been studied with the consideration of spatial reuse which is a potential advantage of D2D networks to further reduce packetizations without reducing per user throughput.

In this paper, we study several approaches to design coded caching networks with reduced packetizations. We propose two combinatorial designs for centralized D2D caching networks which have reduced packetization compared to \cite{ji2016fundamental}. The first approach uses a hypercube to define the cache placement and we demonstrate how the geometry of this hypercube relates to coded multicasting opportunities for delivery. The hypercube approach is optimized specifically for D2D caching networks as opposed to adapting an already studied shared link scheme. In addition, by adopting the idea recently proposed in \cite{jin2016new}, we extend this approach to a decentralized coded D2D caching scheme, which allows a much more flexible design for given network parameters. Meanwhile, the advantage of the reduced packetization of the hypercube approach still remains in the decentralized D2D caching networks. %In the second approach, a
The second approach is based on an application of the Ruzsa-Szem\'eredi graph \cite{ruzsa1978triple, alon2012nearly}, which is first used for shared link caching in \cite{shanmugam2017coded}. %is used to design the placement and delivery phase. The use of Ruzsa-Szem\'eredi graphs for caching network design was first outlined in \cite{shanmugam2017coded} for a shared link scheme.
We extend the use of Ruzsa-Szem\'eredi graph to D2D caching networks. Both D2D combinatorial designs, %without sacrificing much per user throughput of the state-of-the-art schemes \cite{ji2016fundamental},
sustain the significant throughput gain compared to conventional uncoded unicast \cite{Ji2016WirelessD2D} and the required packetizations are reduced exponentially compared to \cite{ji2016fundamental} with respect to the number of users $n$ while keeping the library size $m$ and memory size $M$ fixed. %Finally, we show that both approaches yield a significant reduction in the number file partitions as compared to \cite{ji2016fundamental} while sustaining the significant throughput gain.
Finally, we study the impact of enabling spatial reuse in these caching network designs and show this can further reduce the required packetizations, while also improving the per user throughput significantly for some parameter regimes.

The outline of this paper is as follows. In section \ref{sec: Network Model and Problem Formulation}, we introduce the D2D network model and problem formulation. In Section \ref{sec: Hypercube Caching Network Approach}, we present the hypercube based coded caching approach and analyze its performance. We also show that  this  centralized scheme can be used to design a decentralized D2D caching network.
Section \ref{sec: RS graphs} introduces the Ruzsa-Szem\'eredi graph based coded caching approach and analyzes its performance. In Section \ref{sec: SR},
%we propose the two novel approaches to centralized D2D caching network design, compare the results to the current state-of-the-art design. In section \ref{}
we show how the proposed schemes can take advantage of spatial reuse. Finally, we conclude the paper in Section \ref{sec: Conclusion}. %and in section \ref{} we discuss how the centralized schemes can be used to design a decentralized D2D caching network. Finally, we make closing remarks in section \ref{}.

\section{Network Model and Problem Formulation}
\label{sec: Network Model and Problem Formulation}

\begin{figure}
\centering

\centering \includegraphics[width=10cm, height=5.6667cm]{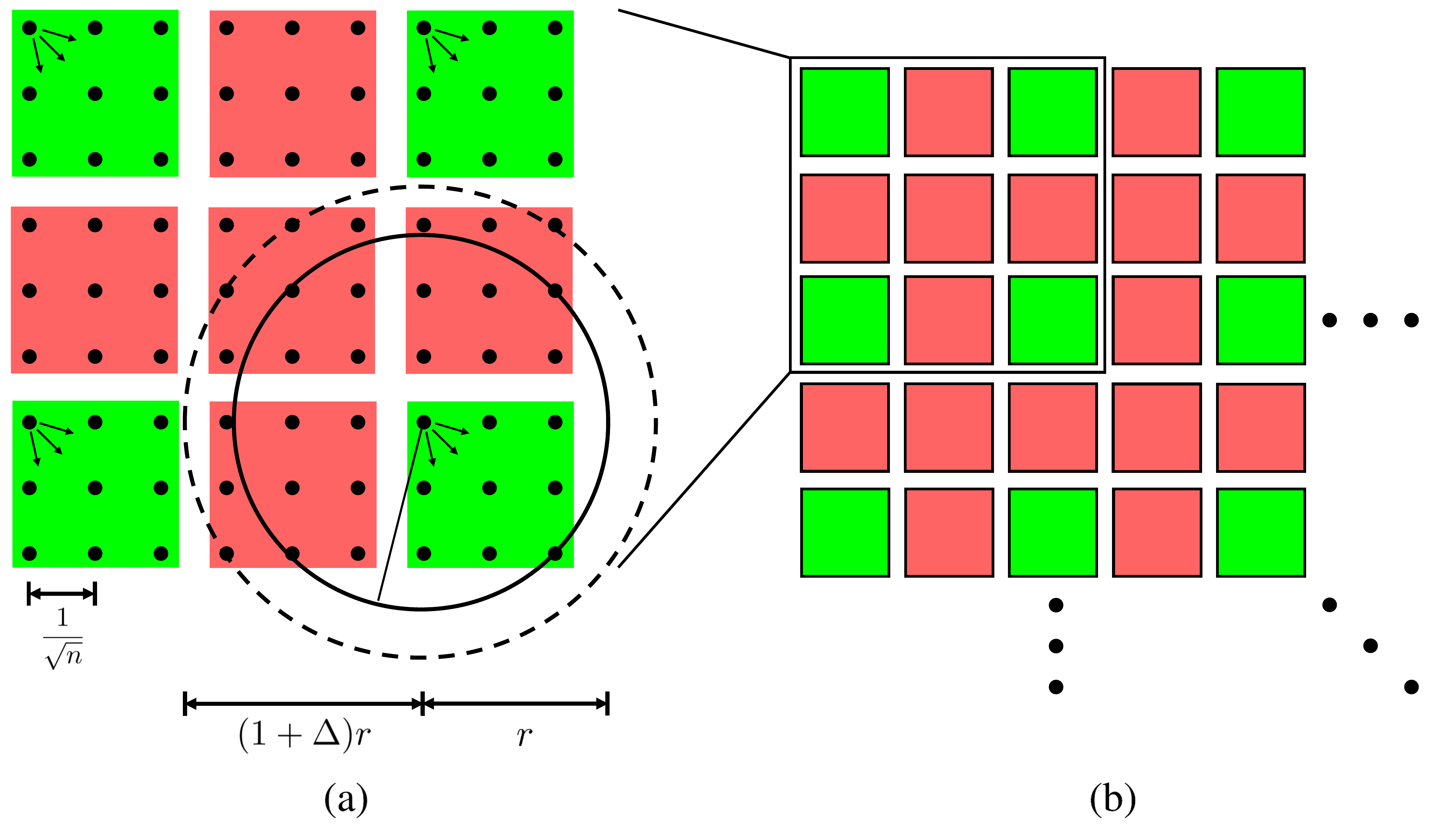} %[width=10cm, height=5.6667cm]

\vspace{-0.2cm}
\caption{~\small a) Representation of 81 users on a square grid. Users are divided into 9 equally sized clusters of 9 users based on user proximity. The clusters highlighted in green represent clusters which can be simultaneously active assuming that cluster highlighted in red are not active. b) A larger network which includes the 81 users of Fig.~\ref{fig: GridNetwork}(a) where the clusters highlighted in green can be simultaneously active. The reuse factor is $\mathcal{K}=4$.%is represented by $3$ data sets. The strong edge coloring is shown by the colors of each edge and the corresponding chromatic index $\chi_{\text{sq}}(\Gc)$ in this graph is $3$.
}
\label{fig: GridNetwork}
\vspace{-0.4cm}
\end{figure}

We consider a wireless D2D network with single-hop transmissions formed by the set of users $\mathcal{U}=%\{ u_0, \cdots, u_{n-1}\} =
\{0, \ldots, n-1\}$. The users are uniformly distributed on a unit grid with a minimum distance of $1/\sqrt{n}$ as shown in Fig.~\ref{fig: GridNetwork}(a). Each user $u\in\mathcal{U}$ makes a request $f_u\in\mathcal{F}$, where $\mathcal{F} = \{1, \cdots, m\}$ is a file library of $m$ independently generated messages $\{ W_0,..., W_{m-1} \}$ with entropy $F$ bits each. We denote the demand vector as $\fsf = (f_0, \cdots, f_{n-1})$. The file library, $\Fc$, is generated once and kept unchanged during subsequent network operations. In addition, we assume $m \geq n$ and users request distinct files.
Each user locally caches the equivalent of $M$ files, or $MF$ bits.
Furthermore, define $t \eqdef nM/m \geq 1$, as the number of times the library is cached collectively among the users.

Users have active links between one another based on the {\em protocol model} \cite{xue2006scaling} described as follows. A communication link, consisting of user $u$ transmitting to user $v$, will be successful if and only if the distance between user $u$ and $v$ is less than or equal to $r$ and user $v$ is at least a distance of $(1+\Delta)r$ from all transmitting users other than user $u$. The parameters $r,\text{ }\Delta>0$ are given by the protocol model. We assume that any $r>0$ is possible and $r$ dictates a constant data rate, $C_r$, in the unit of bits/s/Hz.\footnote{Note that in practice, $C_r$ can be a function of the transmission range $r$. However, the protocol model does not capture this relationship. } %In this work only single-hop D2D transmission is allowed.

The protocol model allows for spatial reuse as shown in Fig.~\ref{fig: GridNetwork}(a). 4 users, one in each green cluster, are transmitting to 8 local users who are at most a distance of $r$ away. The users receiving the transmission are at least a distance of $(1+\Delta)r$ from the other three transmitting users. The users of Fig.~\ref{fig: GridNetwork}(a) may be part of a larger network of users as shown in Fig.~\ref{fig: GridNetwork}(b) which depicts active clusters %of users which are active
(involved in a successful communication link) in green and  non-active clusters %of non-active users
(neither receiving or transmitting) in red. The set of active clusters highlighted in green is one of $\mathcal{K}$ sets. $\mathcal{K}$ is defined as the reuse factor or the number of cluster sets such that, for any given set, each cluster of that set can be active without interference and the $\mathcal{K}$ cluster sets collectively include all clusters.

A D2D caching scheme consists of three phases: the cache placement phase, the coded delivery phase, and the transmission phase. These are defined as follows.

\begin{defn}
\label{def: Cache placement Phase}
{\bf (Cache Placement Phase)} The cache placement phase maps the file library
$\mathcal{F}$ onto the caches of all users $u \in \mathcal{U}$. Every file in $\mathcal{F}$ is split into $K$ equally sized packets and each user caches $MK$ packets or equivalently $MF$ bits. %(i.e., $M$ files).
For $u \in \Uc$, the function $\phi_u: \FF_2^{mF} \rightarrow \FF_2^{MF}$
generates  cache content $Z_u \triangleq \phi_u(W_f :  f \in \Fc)$, which are
%The cache messages $Z_u$
stored in user caches at the beginning,
and kept fixed throughout  subsequent %network
operations.
\hfill $\lozenge$
\end{defn}

\begin{defn}
\label{def: Coded Delivery Phase}
{\bf (Coded Delivery Phase)} The coded delivery phase is defined by two sets of functions:
the node encoding functions, denoted by $\{\psi_u: u \in \Uc\}$, and the node decoding functions,
denoted by $\{\lambda_u: u \in \Uc\}$.  Let $R_u^{\rm T}$ denote the number of {\em coded bits}
transmitted by node $u$ to satisfy the demand vector $\fsf$.
The transmission rate of node $u$ is defined by $R_u = \frac{R_u^{\rm T}}{F}$.
The function $\psi_{u}: \FF_2^{MF} \times \Fc^n \rightarrow \FF_2^{FR_u}$ generates the transmitted message
$X_{u,\fsf}  \triangleq \psi_{u}(Z_u,\fsf)$ of node $u$ as a function of its cache content $Z_u$ and the demand
vector $\fsf$.\footnote{We also refer the transmission rate to traffic load in this paper.}

Let $\Dc_u$ denote the set of users whose transmit messages are received by user $u$
(according to some transmission policy in Definition \ref{def:txpolicy}).
The function $\lambda_u : \FF_2^{F \sum_{v \in \mathcal{D}_u} R_v} \times \FF_2^{MF} \times \Fc^n \rightarrow \FF_2^{F}$
decodes the request of user $u$ from the received messages and its own cache, i.e., we have
\be
\hat{W}_{u,\fsf} \triangleq \lambda_u(\{ X_{v,\fsf} : v \in \Dc_u\}, Z_u, \fsf). \nonumber
\ee
\hfill $\lozenge$
\end{defn}

Since users make arbitrary requests, similar to \cite{maddah2014fundamental,ji2016fundamental}, we focus on the worst-case error probability defined as
\be
\label{eq: error prob}
P_e = \max_{\fsf \in \Fc^n} \; \max_{u \in \Uc} \; \PP\left( \hat{W}_{u,\fsf}  \neq W_{f_u} \right). \nonumber
\ee
For a given number of users $n$ and library size $m$,
letting the transmission rate $R = \sum_{u \in \Uc} R_u$,
we say that the cache-rate pair $(M, R)$ is achievable if $\forall$ $\varepsilon > 0$
there exists a sequence indexed by the file size
$F \rightarrow \infty$ of cache encoding functions $\{\phi_u\}$,
delivery functions $\{\psi_u\}$ and decoding functions $\{\lambda_u\}$, with rate $R^{(F)}$
and probability of error $P_e^{(F)}$ such that $\limsup_{F \rightarrow \infty} R^{(F)} \leq R$ and  $\limsup_{F \rightarrow \infty} P_e^{(F)} \leq \varepsilon$.

Note that $RF$ gives the achievable total traffic load transmitted in the whole network.

\begin{defn} \label{def:txpolicy}
{\bf (Transmission Phase)} The transmission policy $\Pi$
is a rule to activate  D2D links in the network. Let $\Lc$ denote the set of all directed links.
Let $\Ac \subseteq 2^\Lc$ denote the set of all possible feasible subsets of links
(this is a subset of the power set of $\Lc$, formed by all sets of links forming independent sets in the
network interference graph induced by the protocol model).
Let $\Asf_{\tsf} \subset \Ac$ denote a feasible set of simultaneously active links at time $\tsf$.
A feasible transmission policy $\Pi$ consists of a sequence of activation sets, i.e., sets of active transmission links, $\{\Asf_\tsf : \tsf = 1, 2, 3, \ldots\}$, such that at each time $\tsf$ the active links in $\Asf_\tsf$ do not violate the protocol model.
\hfill $\lozenge$
\end{defn}

Different from the shared link network model \cite{maddah2014fundamental}, the performance of D2D caching networks cannot be completely characterized by the transmission rate, $R$, because of spatial reuse under the protocol model.
Hence, we define the per user throughput as follows,
\be
\label{eq: throughput}
T \eqdef \frac{F}{D},
\ee
where $D$ is the number of channel uses required to satisfy all user requests. We say that the pair $(M, T)$  is achievable if $RF$ is achievable and there exists a transmission policy $\Pi$  such that the $RF$ encoded bits can be delivered to their destinations
in $D \leq F/T$ channel uses.
Then, the optimal achievable throughput is defined as
\be
T^*(M) \triangleq \sup\{T : (M, T) \text{ is achievable}\}. \nonumber
\ee

\begin{remark}
\label{remark 1}
Note that, when considering clustering and assuming that the transmission rate of each cluster is exactly $R_{\rm c}$,
we can obtain
$D=\frac{R_{\rm c}F\mathcal{K}}{C_r}$.
Therefore, the per user throughput for a clustering scheme is
\be
\label{eq: Tc}
T_{\rm c} = \frac{C_r}{R_{\rm c}\mathcal{K}}.
\ee
In the following, we will compare the per user throughput of clustering and non-clustering schemes. %using $T_{\rm c}$.
%$\hfill\square$
\end{remark}

\section{Hypercube Coded Caching Approach}
\label{sec: Hypercube Caching Network Approach}

In this section, we consider the case when the transmission range $r \geq \sqrt{2}$. In other words, a transmission from any node %in the network // successfully decoded
can be received by the rest of the nodes in the network and there is no user clustering. We first introduce two motivating examples of the proposed hypercube coded caching approach and then present the general achievable scheme and transmission rate.
%we describe the proposed caching placement and delivery scheme using the hypercube approach. For the ease of illustration, in the next section, we consider a top example.

% every file in the library is split into packets which are represented by points on a hypercube lattice with $t$ dimensions. Each dimension of the hypercube is $m/M$ in length. Each user caches a $t-1$ dimensional plane of this hypercube and forms multicasting groups with users who have cached orthogonal planes. The following theorem presents the achievable  rate and packetization for this scheme.
%\begin{theorem}
%\label{theorem: 1}
%Let $m,n,M$ be the library size, number of users and the cache size per user, respectively. For $n = (m/M)^2$, $m/M \in \mathbb{Z}^+$and $t \geq 2$, the following rate, packetization pair  is achievable:
%\be
%\label{eq: theorem 1}
%(R,K) = (\frac{m}{M},\sqrt{n}^{\sqrt{n}}).
%\ee
%%Moreover, when $t$ is not an integer, the convex lower envelope of $R(M)$, seen as a function of $M \in [0:m]$, is achievable.
%\hfill  $\square$
%\end{theorem}
%\begin{IEEEproof}
%Theorem \ref{theorem: 1} is proved in Sections \ref{sec: Correctness of Scheme} and \ref{sec: Achievable Rate and Packetization}.
%\end{IEEEproof}

\subsection{2-Dimension Example}
\label{sec: 2D example}

\begin{figure}
\centering
\subfigure[]{
\centering \includegraphics[width=7.9cm]{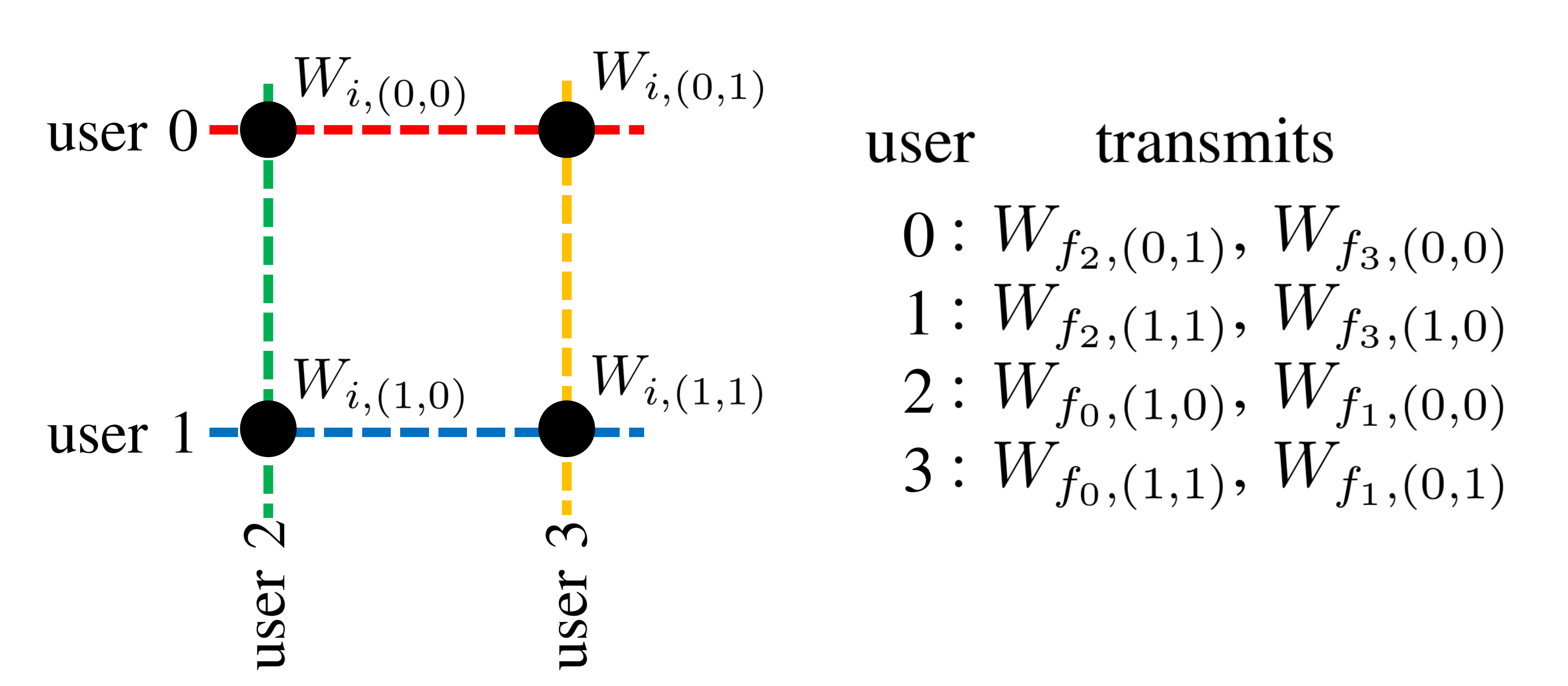} %[width=10cm, height=4.3751cm]{2d_fig}
\label{fig: 2d fig}
}
\subfigure[]{
\centering \includegraphics[width=7.9cm]{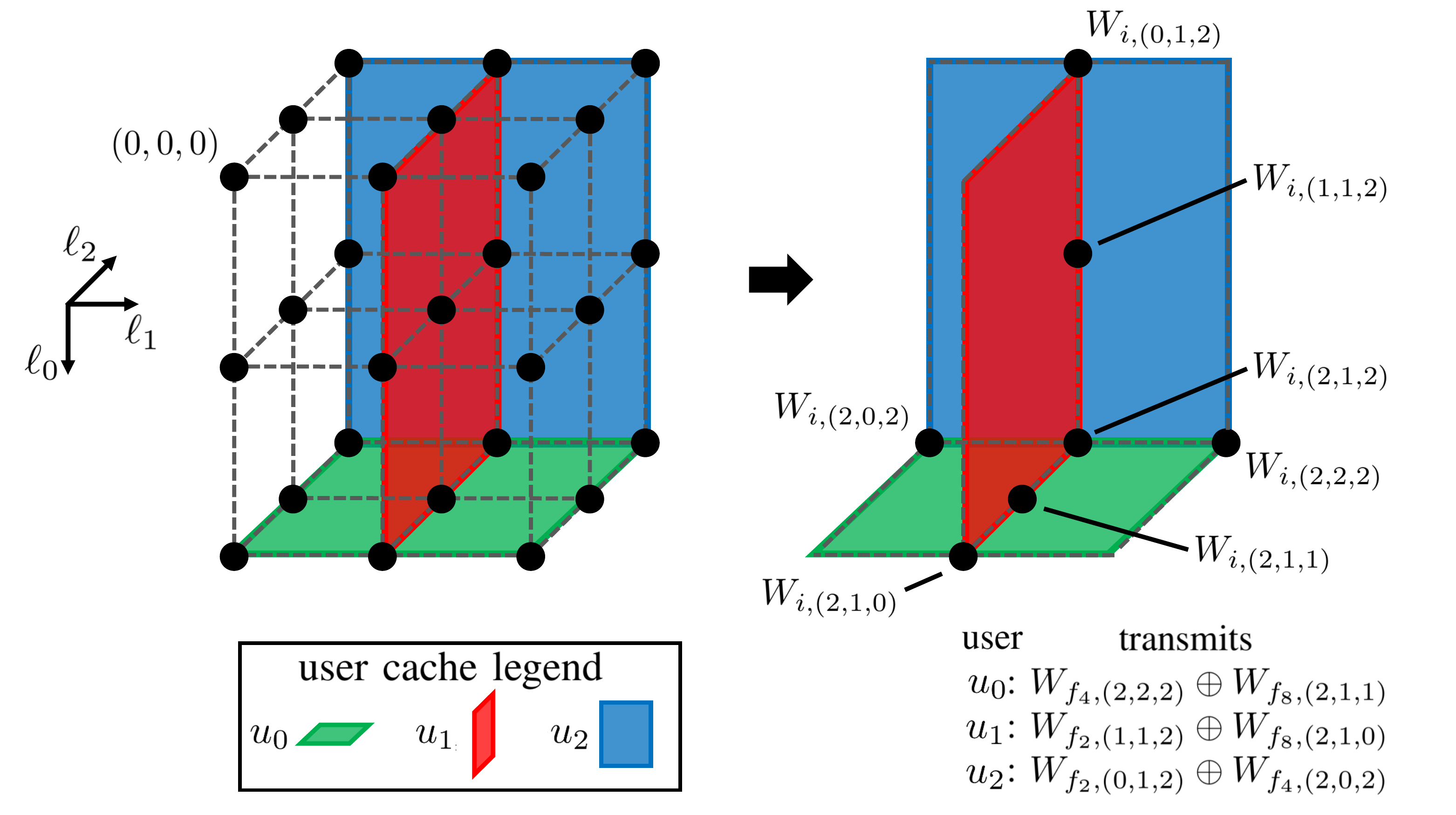}
\label{fig: 3d fig}
}
\vspace{-0.2cm}
\caption{~\small a) An example of the proposed scheme in a D2D network with $n=4$, $M/m=1/2$ and $K=4$. Each point on the lattice represents a set of packets. Each user's cache is represented by a line, where the user caches the points on the lattice which intersect that line. b) Each point on the 3D-lattice represents a set of packets. Each user's cache is represented by a plane of lattice points. The right figure represents a multicasting group, or a collection of $3$ planes, where any two are orthogonal. The points on the intersection of two or more of these planes are highlighted. Every two users in this group share two packets that are requested by the third user. %is represented by $3$ data sets. The strong edge coloring is shown by the colors of each edge and the corresponding chromatic index $\chi_{\text{sq}}(\Gc)$ in this graph is $3$.
}
\vspace{-0.4cm}
\end{figure}

In this example, we propose a cache placement and a delivery scheme based on a $d=2$ dimensional hypercube (a plane) lattice. We consider a D2D network with $n=4$ users, $\frac{m}{M}=2$, and $t=\frac{nM}{m}=2$. %{\RED (I think that we need to introduce $t$ here.)}
Assume that each user requests a distinct file, i.e., user $i$ requests file $f_i$.  %where users cache half the library.
%{\RED (I added some details in the  blue section below to make the example clear. Please check.)}
{The $2$-dimensional lattice is constructed where the size of each dimension is $\frac{m}{M}=2$. Each lattice point represents a set of $m$ packets, each from a different file in $\Fc$. For instance, in Fig.~\ref{fig: 2d fig}, each lattice point represents $\{W_{i, (j,k)}, i \in \Fc\}$ %$\{W_{i, (j,k)}, i=1, \cdots, m\}$,
where $W_{i, (j,k)}$ is a packet from  file $i \in \Fc$ and the subscript $(j,k)$ represents the corresponding position on the lattice. Hence, the total number of lattice points equals the number of packets per file. Since in this case we have $t=d=2$, there are $K=(\frac{m}{M})^t=(\frac{m}{M})^d= 2^2=4$ points in this 2-dimensional lattice, corresponding to $4$ partitions of each file. The number of users equals $t$ times the number of lattice points on each dimension, i.e., $n=\frac{m}{M}t =\frac{m}{M}d =2\cdot 2=4$.}

For the cache placement, each user caches a line on the lattice as shown in Fig.~\ref{fig: 2d fig}. For instance, user 0 caches $W_{i, (0,0)}$ and $W_{i, (0,1)}$ for all $i \in \Fc$. The intersection of two lines represents a set of packets cached by two users. Each user requests two packets (from its demanded file) each of which is in the cache of two other users. In the delivery phase, each user unicasts two requested packets to the appropriate users.
%{\RED (Do we want to introduce user groups here? Check if the newly added blue section helps to explain the transmitted packets shown in  Fig.~\ref{fig: 2d fig} better.) }
{In this example, the number of user groups %(multicasting groups for $t > 2$)
is equivalent to the number of packets per file and there are $K=(\frac{m}{M})^t=2^2=4$ user groups, each with a size of $t=2$, i.e., we pick one user from each dimension. User 0 is involved in two user groups, $\{0,2\}$ and $\{0,3\}$. Hence, user 0 will send $W_{f_2,(0,1)}$ to user 2 and send  $W_{f_3,(0,0)}$ to user 3. }
Therefore, $R=2$ since the rate simply equals the number of packets that the users are missing normalized by the file size. %which is $R=2$.
Note that, in this simple example there is a coded multicasting opportunity. Each user can include their transmitted packets into a single coded multicast. For instance, user $0$ can transmit $W_{f_3, (0,0)} \oplus W_{f_2, (0,1)}$ to users $2$ and $3$, who can decode $W_{f_2, (0,1)}$ and $W_{f_3, (0,0)}$ successfully (in fact, this is precisely the scheme in \cite{wang2017placement} for $t=2$). However, as we will see later,  this pattern will not hold for larger values of $n$ and $t$. The multicasting opportunities which scale with $n$ and $t$  consist of users transmitting requested packets to  a set of $t-1$ users whose cache ``planes'' are aligned along different dimensions. %However, for this simple example, there is no coded multicasting because there are only 2 dimensions of the lattice. The rate is simply equal to the number of equivalent files that the users are missing which is $R=2$. The 3D example will demonstrate how coded multicasting is used to effectively reduce the rate.
This will be shown in the following example.

%\begin{figure}
%\centering
%%\subfigure[]{
%\centering \includegraphics[width=9cm]{2d_fig} %[width=10cm, height=4.3751cm]{2d_fig}
%%}
%%\subfigure[]{
%%\centering \includegraphics[width=7cm, height=5.1cm]{Strong_Edge_Coloring_v2}
%%\label{fig: Strong_Edge_Coloring}
%%}
%\vspace{-0.2cm}
%\caption{~\small An example of the proposed scheme in a D2D network with $n=4$, $M/m=1/2$ and $K=4$. Each point on the lattice represents a set of packets. Each user's cache is represented by a line, where the user caches the points on the lattice which intersect that line. %is represented by $3$ data sets. The strong edge coloring is shown by the colors of each edge and the corresponding chromatic index $\chi_{\text{sq}}(\Gc)$ in this graph is $3$.
%}
%\label{fig: 2d fig}
%\vspace{-0.4cm}
%\end{figure}

\subsection{3-Dimension Example}
\label{sec: 3D example}

We consider a 3-dimensional hypercube (cube in this case), where $d = t = \frac{Mn}{m} = 3$ and each dimension consists of  $\frac{m}{M} = 3$ lattice points. Hence, the number of packets per file is $K=\left(\frac{m}{M}\right)^t=27$.
%$M/m=1/t=1/3$,
Let $n=\frac{m}{M} t=\left(\frac{m}{M}\right)^2 =9$ and user $i$ request file $i$. % with the assumption that the library size $m \geq n$.
%In addition, we denote users $u_0 = 2$, $u_1 = 4$ and $u_2 = 8$.
Similar to the 2-dimensional lattice,  each lattice point on the cube represents a set of $m$ packets, each from a different $W_i$, $i \in \Fc$ %where $i=1, \cdots, m$
(see Fig.~\ref{fig: 3d fig}). For the cache placement, each user caches all packet sets represented by a plane of lattice points of the cube. %by caching all packet sets associated with points where one dimension is fixed and the other two dimensions vary.
%Before introducing the delivery scheme, we will present some important observations of this cache placement scheme.

%\begin{figure}
%\centering
%\centering \includegraphics[width=10cm]{3d_fig_sized_v2} %width=15cm, height=8.4cm
%\caption{~\small Each point on the 3D-lattice represents a set of packets. Each user's cache is represented by a plane of lattice points. The right figure represents a multicasting group, or a collection of $3$ planes, where any two are orthogonal. The points on the intersection of two or more of these planes are highlighted. Every two users in this group share two packets that are requested by the third user. %is represented by $3$ data sets. The strong edge coloring is shown by the colors of each edge and the corresponding chromatic index $\chi_{\text{sq}}(\Gc)$ in this graph is $3$.
%}
%\label{fig: 3d fig}
%\vspace{-0.4cm}
%\end{figure}

In Fig.~\ref{fig: 3d fig}, we show the cache of 3 users, $u_0, u_1$ and $u_2$, depicted by the red, green and blue planes which are orthogonal (let users $u_0 = 2$, $u_1 = 4$ and $u_2 = 8$). %Note that this is a 3-tuple of orthogonal planes.
Given this 3-tuple of orthogonal planes, first consider the packet set $\{W_{i,(2,1,2)}, \forall i \in \Fc\}$ %is not considered because this
which lies on the intersection of all three planes and therefore cached by all 3 users. Any packet from this set does not need to be transmitted among these 3 users. Next, consider packets which are cached by exactly 2 of the 3 users. This is analogous to the points which lie on the intersection of exactly 2 out of the 3 planes. The intersection of the red and green planes yields the points representing the packets $W_{i,(2,1,0)}$ and $W_{i,(2,1,1)}$ for all $i\in\mathcal{F}$. These packets are cached at both users $u_0$ and $u_1$, but not user $u_2$.  %It can be seen that user $u_8$ requests $W_{f_8,(2,1,0)}$ and $W_{f_8,(2,1,1)}$ for which $u_2$ and $u_4$ can each transmit one packet.
It can be seen that $W_{f_8,(2,1,0)}$ and $W_{f_8,(2,1,1)}$ are requested by user $u_2=8$, and are available at user $u_0$ and $u_1$. %Similarly,
%A similar argument will show that
In general,
each user among this 3-tuple requests two packets which are cached at the other two users. %Alternatively, each user among this 3-tuple also caches $4$
Equivalently, this shows a coding opportunity for multicasting from one user to the other two users within this 3-tuple, which will be shown in the following.
%This presents a coded multicasting opportunity where each user can transmit requested packets to all other users by using the XOR function to combine requests. Each user can recover requested content since the coded multicast will contain packets which are either locally cached or requested.

In this delivery phase, each user will multicast the coded packets based on the observations discussed above. In particular, every $3$-tuple of orthogonal planes
%(also representing the corresponding users cache the packets on those planes)
on the cube (representing the cache content of 3 users) represents a coded multicasting group, where each user multicasts one codeword to the other two users. For example, in Fig. \ref{fig: 3d fig}, for the depicted 3-tuple of planes (red, blue, green), user $u_0$ transmits $W_{f_4, (2,2,2)} \oplus W_{f_8, (2,1,1)}$ to $u_1$ and $u_2$; user $u_1$ transmits $W_{f_2, (1,1,2)} \oplus W_{f_8, (2,1,0)}$ to $u_0$ and $u_2$; user $u_2$ transmits $W_{f_2, (0,1,2)} \oplus W_{f_4, (2,0,2)}$ to $u_0$ and $u_1$. By using the cached packets, each user within this multicast group can decode the requested packets. In this case, $3$ coded multicasting packets are transmitted and will result in a traffic load of %an amount of channel use equivalent to
$\frac{3}{27} = \frac{1}{9}$. %files.
%To perform the delivery phase, consider all 3-tuples of planes such that each plane is orthogonal to the other two planes. There are $(m/M)^t = 27$ such sets. Within these sets, the intersection of 2 planes forms a line of lattice points representing a set of packets that two users have cached. Furthermore, this line is orthogonal to the third plane. The third user will request packets associated with this line, with the exception of the point where the third user's plane intersects this line.
%To recover the entirety of a requested file,
%To obtain the requested file,
In general, each user needs packets which are associated with all lines orthogonal to the plane representing that user's cache. These lines are formed by the intersections of all planes which are orthogonal to each other and orthogonal to the plane representing the user's cache. In other words, each user will receive all requested packets by forming multicasting groups with all pairs of users for which all three users cache a plane orthogonal to each other. %Each 3-tuple will yield a transmission rate of $1/9$ and
%Considering all relevant 3-tuples,
{The total number of 3-tuples, or multicasting groups, is equivalent to the number of lattice points, $K$. The transmission rate is equal %equivalent
to the product of the number of multicasting groups, $K=\left( \frac{m}{M}\right) ^t=27$, the number of transmissions per group, $t=3$, and the size of each transmission, $1/\left( \frac{m}{M}\right)^t=\frac{1}{27}$. Therefore, $R=3$.}

\subsection{General Achievable Scheme {for $n={(\frac{m}{M}})^2$}} % $t=\frac{m}{M}$ or
\label{sec: hc general scheme}

We build a hypercube lattice with $t$ dimensions and each dimension has $\frac{m}{M}$ lattice points in length.
%In this scheme, the packets form a hypercube lattice with $t$ dimensions and each dimension is $m/M$ in length.
Each lattice point represents a set of $m$ packets, each from a different $i\in \Fc$. %, where $i \in \Fc$ .%$i=1, \cdots, m$.  %a set of packets.
Each user caches packets represented by the lattice points of a distinct $(t-1)$-dimensional hyperplane of the hypercube by holding one dimension constant and allowing other dimensions to vary. %The cache placement phase is not based on the set of user requests.
In the delivery phase, each multicast group
%the user sets, or multicast groups,
comprises of users, each of which caches a hyperplane that is ``orthogonal'' to the cached hyperplanes of every other user in the group. Within these groups, users multicast a set of coded packets requested by other users in the group. Formally, the cache placement and delivery phase are as follows:

\begin{itemize}

\item {\bf Cache Placement Phase:}
For all $W_i \in \Fc$, partition file $W_i$ into $\left(\frac{m}{M}\right)^t$ packets of equal size labeled as $W_{i,(\ell_0,\ldots ,\ell_{t-1})}$ where $\ell_j \in \{0,\ldots ,\frac{m}{M} -1\}$ for all $j \in \{0,\ldots ,t-1\}.$  Let user $u \in \Uc$ %$u\in \{ 0,\ldots , n-1 \}$
cache a set of packets   $\{W_{i,(\ell_0,\ldots ,\ell_{t-1})}, i=1, \cdots, m\}$ such that $\ell _j = (u \bmod \frac{m}{M})$, where $j=\lfloor u/\frac{m}{M} \rfloor$, and $\ell_i\in \{0,\ldots ,\frac{m}{M} -1\}$ for any  $i \ne j$.
{Hence, each user caches $(\frac{m}{M})^{t-1}$} packets from each file $i \in \Fc$.\footnote{Throughout this paper, we use the following definition for the modulo operation: $a\bmod b = a - \lfloor a/b \rfloor b$. It follows that if $a\in \mathbb{Z}$ and $b \in \mathbb{Z}^+$, then $a \bmod b$ is a non-negative integer in the range of $[ 0, b-1]$.}

\item {\bf Delivery Phase:}
Consider all sets of users $\mathcal{S} = \{ u_0,\ldots , u_{t-1} \} \subset \{ 0,\ldots , n-1 \}$ such that $\lfloor u_j/\left(\frac{m}{M}\right) \rfloor = j$ for all $j \in \{0,\ldots ,t-1\}$ and $|\mathcal{S}|=t$. Each user $u_i \in \mathcal{S}$ transmits
\be
\label{eq: dlv index 1}
\oplus_{s\in \{0,\ldots ,t-1 \} \setminus i} W_{f_{u_s},(a_0,\ldots ,a_{t-1})}
\ee
where
\begin{align}
\label{eq: dlv index 2}
a_q=
\begin{cases}
(u_s + s-i) \bmod \frac{m}{M}, & \text{ if }q=s \\
u_q \bmod \frac{m}{M}, &  \text{ if }q \neq s
\end{cases}
\end{align}

\end{itemize}

The correctness of the proposed scheme is shown in Appendix \ref{sec: Correctness}. To illustrate the notations of the general scheme, we re-state the 3-dimensional example discussed in Section \ref{sec: 3D example}  as follows.

\subsubsection{An Example Using General Scheme for {$n={(\frac{m}{M}})^2$}} %\RED $t=\frac{m}{M}$ or

%To demonstrate the general scheme, we briefly return to the 3D example discussed Section \ref{sec: 3D example} and depicted in Fig.~\ref{fig: 3d fig}.
In Fig.~\ref{fig: 3d fig}, the selected multicast group is a set of three users: $\mathcal{S}=\{2,4,8\}$. We relabel these users so that $u_0=2$, $u_1=4$ and $u_2=8$. This satisfies the condition that $j=\lfloor u_j / \left(\frac{m}{M}\right) \rfloor$ for $j\in \{0,\ldots ,t-1\}$, which ensures that these 3 users cache planes orthogonal to each other. Moreover, $\mathcal{S}\subset \{0,\ldots ,n-1\}$ and $|\mathcal{S}|=t$. Using (\ref{eq: dlv index 1}) and (\ref{eq: dlv index 2}), user $u_0$ (user 2) includes a packet in a multicast for all $s\in \{0,1,2\}\setminus 0$. Those packets are $W_{f_{u_1},(u_0\bmod \frac{m}{M},(u_1+1)\bmod \frac{m}{M},u_2\bmod \frac{m}{M})}$ and $W_{f_{u_2},(u_0\bmod \frac{m}{M},u_1\bmod \frac{m}{M},(u_2+2)\bmod \frac{m}{M})}$. Substituting the values of $u_0$, $u_1$, $u_2$ and $\frac{m}{M}$ yields  packets $W_{f_4,(2,2,2)}$ and $W_{f_8,(2,1,1)}$ which match the packets $u_0$ transmits in Fig.~\ref{fig: 3d fig}.

In the cache placement phase, %since $2 \bmod 3=2$,
user 2 caches all packets $W_{i,(2,\ell_1,\ell_2)}$ for $i\in \{0,...,m-1\}$ and $\ell_1,\ell_2\in\{0,1,2\}$. Both packets transmitted by user 2 satisfy these conditions.
Furthermore,  user 4 only caches packets $\{W_{i,(\ell_0,1,\ell_2)}\}$, and user 8 only caches packets $\{W_{i,(\ell_0,\ell_1,2)}\}$. Hence, we
 recognize that $W_{f_4,(2,2,2)}$ is in user 8's cache, but not in user 4's cache and the opposite is true for $W_{f_8,(2,1,1)}$.
 %Users 4 and 8 cache packets, $W_{i,(\ell_0,\ell_1,\ell_2)}$, only if $\ell_1=1$ and $\ell_2=2$, respectively.
 Therefore, both user 4 and 8 can decode the requested packet as they have the other packet of the coded multicast cached. A similar argument will show the coded multicasts from users 4 and 8 follow  (\ref{eq: dlv index 1}) and (\ref{eq: dlv index 2}) can effectively deliver the requested packets to users $\{2,8\}$ and $\{2,4\}$, respectively.

\subsection{Achievable Rate and Packetization}
\label{sec: Achievable Rate and Packetization}

%Every file in the library is split into packets which are represented by points on a hypercube lattice with $t$ dimensions. Each dimension of the hypercube is $m/M$ in length. Each user caches a $t-1$ dimensional plane of this hypercube and forms multicasting groups with users who have cached orthogonal planes. The following theorem presents the achievable  rate and packetization for this scheme.
The achievable rate using the proposed hypercube based approach is %and packetization pair is
given by the following theorem.
\begin{theorem}
\label{theorem: 1}
Let $m,n,M$ be the library size, number of users and the cache size per user, respectively. For $r \geq \sqrt{2}$, $n = \left(\frac{m}{M}\right)^2$, $\frac{m}{M} \in \mathbb{Z}^+$and $t \geq 2$, the following rate %and packetization pair
is achievable:
%\be
%\label{eq: theorem 1}
%(R,K) = \left(\frac{m}{M},\sqrt{n}^{\sqrt{n}}\right).
%\ee
\be
\label{eq: theorem 1}
R^{\rm hc}(M) = \frac{m}{M}
\ee
with the requirement of $K = K^{\rm hc} = \sqrt{n}^{\sqrt{n}}$.
%Moreover, when $t$ is not an integer, the convex lower envelope of $R(M)$, seen as a function of $M \in [0:m]$, is achievable.
%\hfill  $\square$
\end{theorem}
\begin{IEEEproof}
%Theorem \ref{theorem: 1} is proved in Sections \ref{sec: Correctness of Scheme} and \ref{sec: Achievable Rate and Packetization}.
The number of packets per file is equal to
\be
\label{eq: K}
K^{\rm hc}=\left(\frac{m}{M}\right)^t=\left(\frac{m}{M}\right)^{nM/m}=\sqrt{n}^{\sqrt{n}},
\ee
 or the number of lattice points of each dimension, $\frac{m}{M}$, raised to the power equal to the number of dimensions, $t$.

The multicast groups consist of all possible user sets such that one user is selected from $t$ (number of dimensions) sets of $\frac{m}{M}$ (number of lattice points per dimension) users. Therefore, there are $\left(\frac{m}{M}\right)^t$ multicast groups for which each of $t$ users transmits a codeword with a length of $1/\left(\frac{m}{M}\right)^tF$. The rate is then
\be
R^{\rm hc}(M) = \frac{\left(\frac{m}{M}\right)^{t} \cdot t \cdot  \frac{1}{\left(\frac{m}{M}\right)^{t}}F}{F} %\left(\frac{m}{M}\right)^{t}\left(\frac{M}{m}\right)^{t}t=t=\frac{nM}{m}
=t=\frac{nM}{m}=\frac{m}{M}
\ee
where $n=\left(\frac{m}{M}\right)^2$.
\end{IEEEproof}

The throughput achieved by the proposed achievable scheme is given as follows.
\begin{corollary}
\label{corollary: 11}
Let $C_r$ be the constant link rate under the protocol model. For $r \geq \sqrt{2}$, $n = \left(\frac{m}{M}\right)^2$, $\frac{m}{M} \in \mathbb{Z}^+$and $t \geq 2$, the throughput achieved by the proposed scheme is:
\be  \label{cor 1}
T^{\rm hc}(M) = \frac{C_r}{R^{\rm hc}(M)} = \frac{M}{m}C_r
\ee
%\hfill  $\square$
\end{corollary}
\begin{IEEEproof}
Following  (\ref{eq: throughput}), in order to deliver $F R^{\rm hc}(M)$ coded bits without spatial reuse (at most one active link transmitting at any time),
we need $D = FR^{\rm hc}(M)/C_r$ channel uses. Therefore, we obtain (\ref{cor 1}).
\end{IEEEproof}

%Notice the number of distinct planes defined in this way is exactly $$t\frac{m}{M}=\frac{nM}{m}\frac{m}{M}=n$$ or the product of the number of dimensions and the size of each dimension.
%The number of packets per file is equal to $$K=(\frac{m}{M})^t=(\frac{m}{M})^{nM/m}=\sqrt{n}^{\sqrt{n}}$$ or the size of each dimension, $m/M$, raised to the power of the number of dimensions, $t$.
%
%The multicast groups consist of all possible user sets such that one user is selected from $t$ (number of dimensions) sets of $m/M$ (size of each dimension) users. Therefore, there are $(m/M)^t$ multicast groups for which $t$ users each transmit an equivalent of $1/(m/M)^t$ files. The rate is then $$R = (\frac{m}{M})^{t}(\frac{M}{m})^{t}t=t=\frac{nM}{m}=\frac{m}{M}$$ where the $n=(m/M)^2$.

%\subsection{Achievability for $n < \left(\frac{m}{M}\right)^2$ {\RED Extension to $t< \frac{m}{M}$ or $n < \left(\frac{m}{M}\right)^2$ }}
\subsection{Extension of the achievability to $n \le \left(\frac{m}{M}\right)^2$}
\label{sec: hc general scheme extension}

From the construction of the hypercube scheme, we see that within each multicasting group, each user $u_s\in\mathcal{S}$ needs to  receive $\frac{m}{M}-1$ packets from $t-1$ users. When $n=\left(\frac{m}{M}\right)^2$, we have $t-1=\frac{m}{M}-1$, meaning that each of the $t-1$ users in $\mathcal{S}\setminus u_s$  sends $c=1$ packet to user $u_s$. This guarantees a symmetric multicasting group, where every user transmits the same number of coded packets. The work of \cite{ji2016fundamental} demonstrated that a caching scheme with symmetric multicasting groups generally results in an overall smaller transmission rate than that of a caching network with asymmetric multicasting groups.
%A symmetric multicasting group has been shown to have a smaller transmission rate compared to a non-symmetric multicasting group \cite{ji2016fundamental}.

A more general %requirement which {\RED (A condition that)}
condition that makes the multicasting groups symmetric is
\be
c(t-1)=\frac{m}{M}-1,
\ee
where $c\in\mathbb{Z}^+$. In this case, for a multicasting group, each user requests $c(t-1)$ packets that the rest of the $t-1$ users have cached. Each user transmits $c$ packets that a user requests. Note that these $c$ packets cannot be included in a single coded multicast because the receiving user cannot decode more than one unknown packet and therefore cannot recover the requested packets. Hence, for each multicasting group, each user will transmit $c$ coded multicast messages. In this case, %{\RED (in the following, consider replacing $K_{\rm g}^{\rm hc}$ with $K_{\rm s}^{\rm hc}$ to emphasize symmetric condition rather than general condition.)}
the required number of packets per file $K = K_{\rm g}^{\rm hc}=\left(\frac{m}{M}\right)^t$ is identical to $K^{\rm hc}$.
 By substituting $c = (\frac{m}{M}-1)/(t-1)$, the transmission rate is given by
\be
\label{eq: R small n}
R_{\rm g}^{\rm hc}(M) = \left(\frac{m}{M}\right)^t\left(\frac{m}{M}\right)^{-t}ct=\frac{t}{t-1}\frac{m}{M}\left(1-\frac{M}{m}\right).
\ee
{We note that an exact description of the extension to the case of  $n \leq \left( \frac{m}{M}\right)^2$ requires a modification of the delivery phase shown in Section \ref{sec: hc general scheme}  for $n = \left( \frac{m}{M}\right)^2$ and its correctness follows from a modified version of the proof in Appendix \ref{sec: Correctness}. These are omitted due to space limitation.}
%where $c = (m/M-1)/(t-1)$. It
%a_j = mod(k_j/(m/M)) \text{ }\forall \text{ }k_j \in \mathcal{S} \setminus k_s)$ from user

\subsection{Comparison to State-of-the-Art D2D Caching Networks}
\label{sec: comparison 1}

As was proposed in \cite{ji2016fundamental}, the achievable rate and packetization pair is
\be
(R', K')=\left(\frac{m}{M}\left(1-\frac{M}{m}\right), t\binom{n}{t}\right) = \left(\frac{m}{M}-1, t\binom{n}{t}\right) .
\ee
Implementing Stirling's formula demonstrates that
\be
K' = t\binom{n}{t}\sim \sqrt{\frac{M}{2\pi (m-M)}}e^{\ln{n}+n(\frac{M}{m}\ln{\frac{m}{M}}+(1-\frac{M}{m})\ln{\frac{m}{m-M}})},
\ee
as $n\rightarrow\infty$. We are interested by what factor the hypercube approach has reduced the necessary packetizations in comparison to the scheme in \cite{ji2016fundamental}:
\be
\frac{K'}{K_{\rm g}^{\rm hc}}\sim \sqrt{\frac{M}{2\pi (m-M)}}e^{\ln{n}+n(1-\frac{M}{m})\ln{\frac{m}{m-M}}},
\ee
which grows exponentially as $n \rightarrow \infty$.
Furthermore, we are interested in comparing $R'$ and $R$ achieved by the scheme in \cite{ji2016fundamental} and the proposed hypercube based approach. Hence, we obtain
\be
\frac{R'(M)}{R_{\rm g}^{\rm hc}(M)}=\frac{t-1}{t}=1-\frac{m}{nM},
\ee
It is clear that, given a constant $m$ and $M$,
\be
\lim\limits_{n\to \infty} \frac{K'}{K_{\rm g}^{\rm hc}}=\infty,
\ee
and
\be
\lim\limits_{n\to \infty} \frac{R'(M)}{R_{\rm g}^{\rm hc}(M)}=1.
\ee
%In particular, when $n = \left(\frac{m}{M}\right)^2$, the additive gap between
The hypercube approach demonstrates a significant decrease in the number of packetizations, especially as the number of users becomes large. Furthermore, the rate only increases slightly, and for a large number of users the rate of the two schemes are essentially equivalent.

\subsection{Decentralized Coded Caching using Proposed Scheme}
\label{sec: decentralized hypercube}

%Motivated by
Using the idea of %from the work in
\cite{jin2016new}, we will now explore the use of  centralized D2D caching schemes to design decentralized coded D2D caching networks. A decentralized network consists of $n$ users, %and each user
randomly caches a $\frac{M}{m}$ fraction of  content files. In the following, we focus on the case that $r \geq \sqrt{2}$. A centralized cache placement phase defines sets of packets from which users randomly cache one of these sets independently of  other users' cache. Then, a %slightly
modified delivery phase is used to serve the user requests.

More specifically, we consider the file partition and cache placement of the proposed D2D caching scheme designed for $n'$ (dummy) users such that each of them caches a $\frac{M}{m}$ fraction of the content files, where $n' \ll n$.\footnote{Note that %the ``$n$ users" are not part of the $n$ users in the decentralized network. Here,
$n'$ is only a dummy variable to define the % finite number of
packet sets %available
for the $n$ users in the decentralized network to cache.}  This means that we partition each file into $K'=\left(\frac{m}{M}\right)^{t'}$ packets defined by the centralized cache placement proposed in Section \ref{sec: hc general scheme} and \ref{sec: hc general scheme extension}, where $t' = \frac{n'M}{m}$. Then, we have $n'$ sets of packets and each of which is designed to be cached by one of the $n'$ dummy users.
Each of the $n$ users in the decentralized network randomly caches one of the $n'$ packet sets defined by the centralized scheme. The probability of caching any packet set is $\frac{1}{n'}$. The coded multicasting opportunities become clear as we consider subsets of users in the network who have collectively cached all of the $n'$ packet sets exactly once. In this case, these users can have their requests satisfied by performing the corresponding delivery phase defined by the centralized scheme. This process is repeated until the remaining unsatisfied users do not collectively cache all packet sets. These remaining users are served with a slightly modified delivery phase.
%Furthermore, we define $t' \eqdef \frac{n'M}{m}$ and the number of packets per file is equal to $K'$ or the number of packets per file defined by the centralized cache placement phase.
The following example demonstrates this process for the hypercube approach.

\subsubsection{An Example}

\begin{figure}
\centering
%\subfigure[]{
\centering \includegraphics[width=6.5in, height=2.25in]{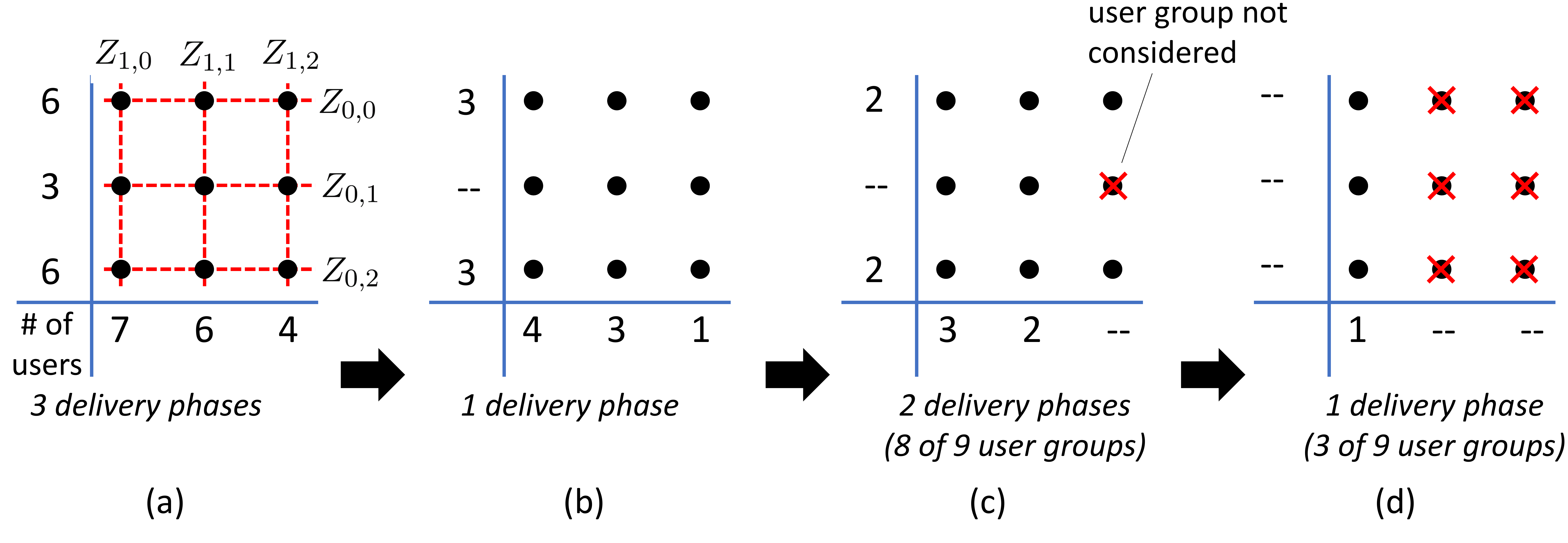}
%}
%\subfigure[]{
%\centering \includegraphics[width=7cm, height=5.1cm]{Strong_Edge_Coloring_v2}
%\label{fig: Strong_Edge_Coloring}
%}
\vspace{-1.2cm}
\caption{~\small A depiction of the caching and delivery phases using the hypercube approach to design decentralized D2D caching networks. The lattice structure represents the file packetization and the numbers to the left and bottom of the lattice are the number of remaining unsatisfied users who cache the rows and columns respectively. The red ``X''s represent pairs of users that are not considered in the delivery phase.
}
\label{fig: dec fig}
\vspace{-0.4cm}
\end{figure}

In this example, a decentralized network of $n=32$ users and each user caches %has the caching capacity to store
$\frac{M}{m}=\frac{1}{3}$ of the library. %per user.
Let $n'=6$ and $t'=\frac{n'M}{m}=2$.
There are $6$ packet sets %available to the users to cache which are
defined by a  2-dimensional hypercube described in Section \ref{sec: hc general scheme extension}. %with $n'=6$ and $t'=\frac{n'M}{m}=2$.
As shown in Fig.~\ref{fig: dec fig}, there are $\frac{m}{M}=3$ lattice points on each dimension of the hypercube.  The number of packets per file for this decentralized network is $K'=K_{\rm g}^{\rm hc}=(\frac{m}{M})^{t'}=9$.  The packet sets, corresponding to each row or column of the hypercube, are labeled as ${\Zc}_{i,j}$ where $i\in \{0,\ldots ,t'-1\}$ and $j\in \{0,\ldots ,\frac{m}{M}-1\}$. %Here, $\Zc_{i,j}$ also represents the set of packets that user $\frac{m}{M}i + j$ would have cached in the centralized scheme.
Each of the $32$ users  caches one of these sets at random using a uniform distribution,  independently of other users,  such that each set has an equal probability of being cached at any given user. A possible outcome of the random caching is shown in Fig.~\ref{fig: dec fig}. Let the random variable $X_{i,j}$ be the number of users that cache $\Zc_{i,j}$ and $x_{i.j}$ be the realization of $X_{i,j}$. In this case, we have $x_{0,0}=6$, $x_{0,1}=3$, $x_{0,2}=6$, $x_{1,0}=7$, $x_{1,1}=6$ and $x_{1,2}=4$. For instance, $6$ users cache the $0$-th row of packets, $\Zc_{0,0}$, and $4$ users cache the $2$nd column of packets, $\Zc_{1,2}$.

To satisfy the $32$ distinct user requests, there are multiple {delivery} phases. First, observe that we can choose $\min\{x_{0,0}, x_{0,1}, x_{0,2}, x_{1,0}, x_{1,1},x_{1,2}\} = 3$ non-overlapping user sets each consisting of $6$ users %for which the users
who collectively cache the $6$ packet sets (see Fig.~\ref{fig: dec fig}(a)). These $3$ user sets can perform the %independent
delivery phases as if they were centralized networks. After this, there are $32-3\cdot 6=14$ remaining users with unsatisfied requests as shown in Fig.~\ref{fig: dec fig}(b). Of these remaining users, we consider a set of $5$ users who collectively cache every packet set except $\Zc_{0,1}$. These $5$ users can perform a centralized delivery scheme by including one previously satisfied user who caches $\Zc_{0,1}$.\footnote{We can simply let a satisfied user request empty packets.} In the third phase, as shown in Fig.~\ref{fig: dec fig}(c), of the remaining $14-1\cdot 5=9$ unsatisfied users, we consider $2$ sets of $4$ users who collectively cache the packet sets $Z_{0,0}$, $\Zc_{0,2}$, $\Zc_{1,0}$ and $\Zc_{1,1}$. These user sets can perform the delivery phases by including previously satisfied users that cache $\Zc_{0,1}$ and $\Zc_{1,2}$. Note that the points of the $3$ by $3$ lattice not only represent a packet for every file in the library, but also represent all multicasting groups in the delivery phase.
In this particular example, there is one multicasting group marked by the red ``X" in Fig.~\ref{fig: dec fig}(c) that does not need to be considered since both users in this multicasting group are satisfied.
%The points of the $3$ by $3$ lattice not only represent a packet for every file in the library, but also represent all multicasting groups in the centralized scheme. Or in other words, each lattice point represents the intersection of $t'$ orthogonal planes, or $2$ lines in this case. The red ``X'' in Fig.~\ref{fig: dec fig} denotes a multicasting group of users who have already been satisfied, and therefore no transmissions from this group are necessary.
%Finally,
Similarly, after the delivery shown in Fig.~\ref{fig: dec fig}(c), %{\BLUE
we are left with a single user (but forming 3 multicast groups) who can obtain the requested packets from the $3$ users caching these packets.  %without forming a multicast group.} {\RED check here. This sentence doesn't match with the caption of Fig. 3d which mentions 3 out of 9 user groups} %who can form multicasting groups as necessary to receive the requested packets.
The transmission rate of %the decentralized rate of
this example is given by
\be
R'=\left(3+1+2 \cdot  \frac{8}{9}+3 \cdot \frac{1}{9}\right) R_{\rm g}^{\rm hc} = \frac{220}{9}. %{\RED \text{ missing multiply by} R_{\rm g}^{\rm hc}=4.}.
\ee

This decentralized scheme does not require any additional packetization. %and the delivery scheme is simple.
Interestingly, note that
%In addition, this scheme is just for the purpose of illustration. It can be observed that
the transmission rate of this scheme is %in fact an uncoded unicast scheme whose transmission rate is
even larger than that of conventional unicasting with  %However, % with transmission rate of
%Also, it is clear this approach slightly overestimates the rate which is clear since $R'$ is greater than the uncoded rate,
$R^{\rm u}=n\left( 1 - \frac{M}{m} \right) = \frac{64}{3}$. This occurs because there is only unicasting for the 2-dimensional hypercube approach %, and therefore $R'$ and $R^{\rm u}$ should be equivalent, except
and we assume every user transmits in every applicable user pair (multicasting group), %for $t'>2$),
which allows unnecessary transmissions as users would be transmitting to users with previously satisfied requests (e.g., in Fig.~\ref{fig: dec fig}(b), we assume users still serve users caching the packet sets of the middle row).
%{\RED I thought in this example we already excluded transmissions to users with previously satisfied requests.}
We want to emphasize that this example is just for the purpose of illustration and this effect on the overestimation of the rate diminishes for a larger values $n$ and $t'$ (see Fig. \ref{fig: sim results}).
%In addition, in this particular example, it is possible to form a multicasting group with $3$ users as discussed in the end of Section \ref{sec: 2D example}. By using this delivery scheme, the achievable transmission rate is $\left(\right)$

\subsubsection{General Decentralized Algorithm}

In this subsection, we describe a general algorithm for using the hypercube scheme for the design of decentralized coded D2D caching networks. The cache placement and delivery phases are defined as follows:

\begin{itemize}

\item {\bf Cache Placement Phase:}
Define $n'\ll n$ such that $t' = \frac{n'M}{m}\in \mathbb{Z}^+$. For all $k \in \Fc$, split file $k$ into $\left(\frac{m}{M}\right)^{t'}$ packets of equal size labeled as $W_{k,(\ell_0,\ldots ,\ell_{t'-1})}$ where $\ell_q \in \{0,\ldots ,\frac{m}{M} -1\}$ for all $q \in \{0,\ldots ,t'-1\}$. For all $i\in \{0,\ldots ,t'-1\}$ and $j\in \{0,\ldots ,\frac{m}{M}-1\}$, define packet set $\Zc_{i,j}$ to include all packets $W_{k,(\ell_0,\ldots ,\ell_{t'-1})}$ such that $\ell_i = j$. Let every user $u\in \{0,\ldots ,n-1 \}$ uniformly caches one of the $n'$ packets sets at random with probability $\frac{1}{n'}$.

\item {\bf Delivery Phase:}
Define $\boldsymbol{X}$ as a $t'\times \frac{m}{M}$ matrix where element $x_{i,j}$ equals the number of users who cache the packet set $\Zc_{i,j}$. While $\boldsymbol{X}$ has non-zero elements, we do the following.
%While $\boldsymbol{X}$ has non-zero elements:
1) let $X_i^\ast$ be the number of zero elements in the $i$-th row of $\boldsymbol{X}$;
 let $\hat{X}^\ast=\sum_{i=0}^{t'-1}X_i^\ast$.
 2) let $x$ be the minimum non-zero element of $\boldsymbol{X}$.
 %{\RED (Should the following be an iterative procedure?, Yes "While $\boldsymbol{X}$ has non-zero elements:" was meant to show it is iterative. Perhaps there is a better way to make it clear?)}
 3) let user sets $\{\mathcal{S}_1,\ldots,\mathcal{S}_x\}$ be %non-overlapping
 disjoint sets each of which contains $n'-\hat{X}^\ast$ unsatisfied users. Each of this user set collectively caches every packet set $\Zc_{i,j}$ that satisfies $x_{i,j} \neq 0$; 4) define user sets $\{\mathcal{S}_1',\ldots,\mathcal{S}_x'\}$ such that each user set has size $\hat{X}^\ast$ and collectively caches every packet set $\Zc_{i,j}$ that satisfies $x_{i,j} =0$. Note that $\Sc_k \cap \Sc_k' = \emptyset$. 5) for each $k\in \{1,\ldots ,x\}$, let users in $\mathcal{S}_k \cup \mathcal{S}_k'$ %{\RED (Can we assume that the two sets are disjoint?, Yes, one group is unsatisfied users, the other group is satisfied users)}
 perform a centralized hypercube D2D delivery phase while ignoring the $\prod_{i=1}^{t'-1}X_i^\ast$ multicasting groups which contain only previously satisfied users; 6) %{\BLUE re-define $\boldsymbol{X}$ such that each element
 update $x_{i,j}$  as $x_{i,j}=\max(0, x_{i,j}-x)$. %}
 7) %{\RED (Should we add this?, Looks good to me)
 go back to step 1) until all elements in $\boldsymbol{X}$ are zero.
 %$\boldsymbol{X}=\boldsymbol{X}-x$ and 6) for all negative %elements $X_{i,j}$ of $\boldsymbol{X}$ set $X_{i,j}=0$.

\end{itemize}

\subsubsection{Performance}
Performance of the decentralized design is given in the following theorem.
\begin{theorem}
Let $m,n,M$ be the library size, number of users and the cache size per user, respectively. Assume that $r \geq \sqrt{2}$, $c(t'-1) = \frac{m}{M}-1$, %$n' = \left(\frac{m}{M}\right)^2$,
where
$(\frac{m}{M}, c) \in \mathbb{Z}^+$,  $t'=\frac{n'M}{m} \geq 2$, and the required number of packets per file $K' = \left(\frac{m}{M}\right)^{t'}$. Furthermore, if $n \geq \beta n'\log n'$ for some $\beta > 1$, then all the packet sets can be cached in the network with probability $1-{n'}^{1-\beta}$, and the following transmission rate is achieved with probability $1-o(1)$ as $n ,n' \rightarrow \infty$, %and packetization pair
%is achievable:
%\be
%\label{eq: theorem 1}
%(R,K) = \left(\frac{m}{M},\sqrt{n}^{\sqrt{n}}\right).
%\ee
\be
\label{eq: theorem decentralized}
R^{\rm hc}_{\rm d}(M) \leq k_\alpha \frac{t'}{t'-1}\frac{m}{M}\left( 1-\frac{M}{m} \right),
\ee
where
\begin{align}
\label{eq: k alpha}
 k_\alpha=
\left\{ \begin{array}{cc} (d_\beta-1+\alpha)\log n'\frac{t'}{t'-1}\frac{m}{M}\left( 1-\frac{M}{m} \right), & {\rm if} \; n=\beta n'\log n', \\
\left(\frac{n}{n'}+\alpha\sqrt{\frac{2n}{n'}\log n'}\right) \frac{t'}{t'-1}\frac{m}{M}\left( 1-\frac{M}{m} \right), &  {\rm if} \;  \omega\left(n'\log n'\right) = n \leq n \cdot {\rm polylog} (n), \\
\left(\frac{n}{n'}+\sqrt{\frac{2n\log n'}{n'}}\left(1 - \frac{1}{\alpha} \frac{\log\log n'}{2\log n'}\right)\right) \frac{t'}{t'-1}\frac{m}{M}\left( 1-\frac{M}{m} \right), &  {\rm if} \; n = \omega\left(n \left(\log n\right)^3\right).
\end{array} \right.
\end{align}
Here, $\alpha>1$, $d_\beta$ is a number depending only on $\beta$ and $ {\rm polylog} (n)$ denotes the class of functions $\bigcup_{k \geq 1} O\left((\log n)^k\right)$.
\end{theorem}
\begin{IEEEproof}
First, we show that with probability $1-{n'}^{1-\beta}$, all the packet sets can be cached in the D2D network. Since each packet set is uniformly selected by each user at random, it can be observed that it is a famous ``bin-ball" problem, where the packet sets are the $n'$ bins and users are the $n$ balls. Hence, we need to characterize the probability that no bin is empty. %In this case, it is also the famous ``coupon collector" problem. Hence, if we
Let $Y_k$ demote the event that the $k$ bin is empty, then we can compute
\be
\PP\left(Y_k\right) = \left(1-\frac{1}{n'}\right)^n \leq e^{-n/n'}.
\ee
Hence
\begin{eqnarray}
\PP\left(\cap_k \overline{Y_k}\right) &=& \PP\left(\overline{\cup_k {Y_k}}\right) = 1 - \PP\left({\cup_k {Y_k}}\right)
\geq 1 - \sum_k  \PP\left({{Y_k}}\right) \geq 1 - n' e^{-n/n'} \notag\\
&\geq&  1 - n' e^{- \beta n'\log n'/n'}  = 1 - {n'}^{1-\beta}.
\end{eqnarray}
Second, we will show (\ref{eq: theorem decentralized}) holds with probability $1-o(1)$. Let $L$ denote the maximum number of users that cache the same packet sets, or equivalently,  the maximum number of balls in each bin.
This problem is to find $k_\alpha$ such that the tail estimate $\PP\left( \left\{L \geq k_\alpha\right\} \cap \left\{\cap_k \overline{Y_k}\right\} \right) = 1 - o(1)$. Hence, by using (\ref{eq: R small n}), we can obtain $R^{\rm hc}_{\rm d}(M) \leq L \frac{t'}{t'-1}\frac{m}{M}\left( 1-\frac{M}{m} \right) \leq k_\alpha \frac{t'}{t'-1}\frac{m}{M}\left( 1-\frac{M}{m} \right)$ with probability $1-o(1)$. From Theorem 1 in \cite{raab1998balls}, we can show that if $k_\alpha$, $\alpha > 1$ satisfies (\ref{eq: k alpha}), then we must have $\PP\left( \left\{L \geq k_\alpha\right\} \right) = 1 - o(1)$. Furthermore,  since
\begin{eqnarray}
\PP\left( \left\{L \geq k_\alpha\right\} \right)  &=& \PP\left( \left\{L \geq k_\alpha\right\} \cap \left\{\cap_k \overline{Y_k}\right\} \right) +  \PP\left( \left\{L \geq k_\alpha\right\} \cap \left\{\cup_k {Y_k}\right\} \right) \notag\\
&=&  \PP\left( \left\{L \geq k_\alpha\right\} \cap \left\{\cap_k \overline{Y_k}\right\} \right) + \PP\left(\left\{\cup_k {Y_k}\right\} \right) \PP\left(\left\{L \geq k_\alpha\right\} |\left\{\cup_k {Y_k}\right\} \right) \notag\\
&\leq& \PP\left( \left\{L \geq k_\alpha\right\} \cap \left\{\cap_k \overline{Y_k}\right\} \right) + {n'}^{1-\beta},
\end{eqnarray}
 we obtain $ \PP\left( \left\{L \geq k_\alpha\right\} \cap \left\{\cap_k \overline{Y_k}\right\} \right) \geq \PP\left( \left\{L \geq k_\alpha\right\} \right) - {n'}^{1-\beta} = 1-o(1)$. This completes the proof.
\end{IEEEproof}

\subsubsection{Simulation Methods and Results}

\begin{figure}
\centering
%\subfigure[]{
\centering \includegraphics[width=6.5in, height=2.1461in]{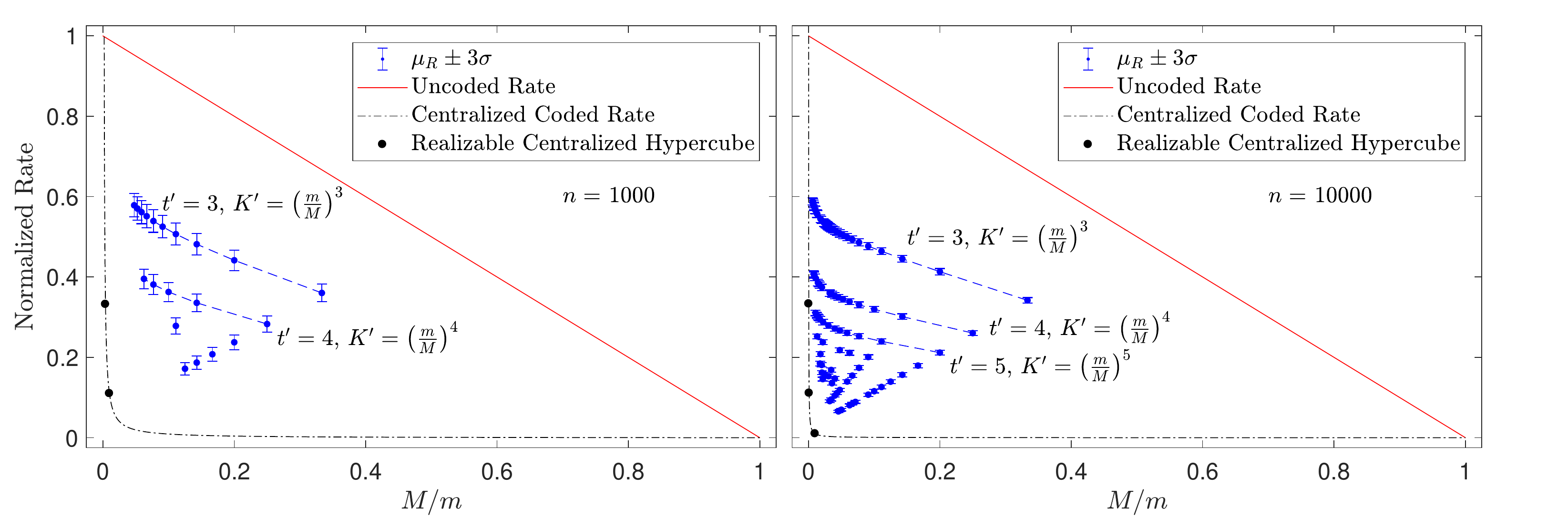}
%}
%\subfigure[]{
%\centering \includegraphics[width=7cm, height=5.1cm]{Strong_Edge_Coloring_v2}
%\label{fig: Strong_Edge_Coloring}
%}
\vspace{-1.2cm}
\caption{~\small Results of the hypercube approach decentralized D2D caching network simulations for $n=1000$ users (left plot) and $n=10000$ users (right plot). The mean rate $\mu_R = \EE\left[R^{\rm hc}_{\rm d}(M)\right]$ plus or minus 3 times the standard deviation, $\mu_R \pm 3 \sigma$, for a variety of hypercube constructions are compared to the uncoded rate and centralized rate. %{\RED need to make the dots bigger for centralized schemes. }
}
\label{fig: sim results}
\vspace{-0.4cm}
\end{figure}

Decentralized networks are simulated with the number of users, $n$, equals to $1000$ and $10000$ users, respectively. For these networks we simulate a decentralized cache scheme for all parameter pairs, $n', t' \in \mathbb{Z}^+$, such that $t' \geq 3$, $n'\leq 64$ for the network of 1000 users and  $n'\leq 500$ for the network of 10000 users, and %the following equation holds
%\be
$c\left( t'-1\right) = \frac{m}{M}-1$ holds,
%\ee
where $c\in \mathbb{Z}^+$ and $t'=\frac{n'M}{m}$. For each parameter pair, $10^4$ decentralized cache placement and delivery phases are simulated. Given that $n'\ll n$ in all cases, there were no instances of a simulated decentralized network where a packet set was not cached at least once. The mean and standard deviation of the rate, $\mu _R$ and $\sigma$, respectively, are depicted in  Fig.~\ref{fig: sim results}.
%$R=3R_{\rm g}^{\rm hc}+R_{\rm g}^{\rm hc}+2\frac{8}{9}R_{\rm g}^{\rm hc}+\frac{3}{9}R_{\rm g}^{\rm hc}$
%$Z_{0,0}$ $Z_{0,1}$ $Z_{0,2}$ $Z_{1,0}$ $Z_{1,1}$ $Z_{1,2}$

From these results, we can see that the decentralized hypercube scheme outperforms  the uncoded scheme. On both plots, the set of points with a fixed $t'$ is highlighted. As $t'$ increases, the rate comes closer to the centralized coded rate. However, the number of packets per file, $K'$, increases exponentially. This exciting result demonstrates that there is a trade-off in designing decentralized D2D caching networks using the hypercube approach. Specifically, we can increase the number of packets to reduce the rate and vice versa. This may provide the flexibility to yield a practical amount of packetization while having limited impact on the transmission rate. Furthermore, we can see that the decentralized approach yields more realizable constructions of the hypercube. In fact, as shown in Fig.~\ref{fig: sim results}, the number of realizable hypercube schemes for 1000 and 10000 users are 2 and 3 respectively ($t=4, 10$ for $n=1000$ and  $t=4, 10, 100$ for $n=10000$). In comparison, the decentralized network design provides many more possibilities.

\section{Ruzsa-Szem\'eredi graph Coded Caching Approach}
\label{sec: RS graphs}

While the hypercube approach requires significantly (in fact, exponentially) less number of packets per file compared to \cite{ji2016fundamental}, the hypercube approach still yields an exponential number of packets relative to $n$ %the number of users
if $m$ and $M$ are fixed. While the hypercube approach certainly increases the domain for which a D2D caching network is implementable, it  is still an open question as to whether there exists a coded D2D caching network scheme without spatial reuse ($r > \sqrt{2}$) with a sub-exponential number of packets per file. Motivated by \cite{shanmugam2017coded}, in this section we propose a coded D2D caching scheme based on  Ruzsa-Szem\'eredi graphs %This is the motivation for our work on Ruzsa-Szem\'eredi graphs,
which requires only a sub-quadratic packetization. While the general, expandable scheme only holds for arbitrarily large $n$, this work demonstrates that sub-quadratic D2D caching schemes exist.

\subsection{Ruzsa-Szem\'eredi Graphs}
%\label{sec: RS graphs}

In this section, we focus on a specific %introduce a
Ruzsa-Szem\'eredi graph design, which  was first introduced in \cite{ruzsa1978triple} and used in a novel manner in \cite{shanmugam2017coded} to construct a cache placement and coded multicasting scheme for a shared link caching network. Let $\Gc(\Vc, \Ec)$ be an undirected graph, where $\Vc$ is the vertex set and $\Ec$ is the edge set. We introduce the following definitions \cite{shanmugam2017coded,alon2012nearly}.
\begin{defn}
Given a graph  $\Gc(\Vc, \Ec)$, a matching $\Mc$ in $\Gc$ is a set of pairwise non-adjacent edges; that is, no two edges share a common vertex.
\hfill $\lozenge$
\end{defn}
\begin{defn}
The set of edges $\Mc \in \Ec$ is an induced matching if for the set $\Sc$ of all the vertices incident on the edges in $\Mc$, the induced graph $\Gc_{\Sc}$ contains no other edges apart from those in the matching $\Mc$.
\hfill $\lozenge$
\end{defn}
\begin{defn}
A disjoint matching $\Mc$ is a matching such that any pair of the edges in $\Mc$ are not adjacent to any third edge in $\Ec$.
\hfill $\lozenge$
\end{defn}
\begin{defn}
A graph $\Gc(\Vc, \Ec)$ is called an $(\gamma,\tau)-$Ruzsa-Szem\'eredi (RS) Graph if its set of edges consists of $\tau$ pairwise disjoint induced matching, each of size $\gamma$.
\hfill $\lozenge$
\end{defn}

\begin{figure}
\centering
\subfigure[]{
\centering \includegraphics[width=7cm, height=5.7cm]{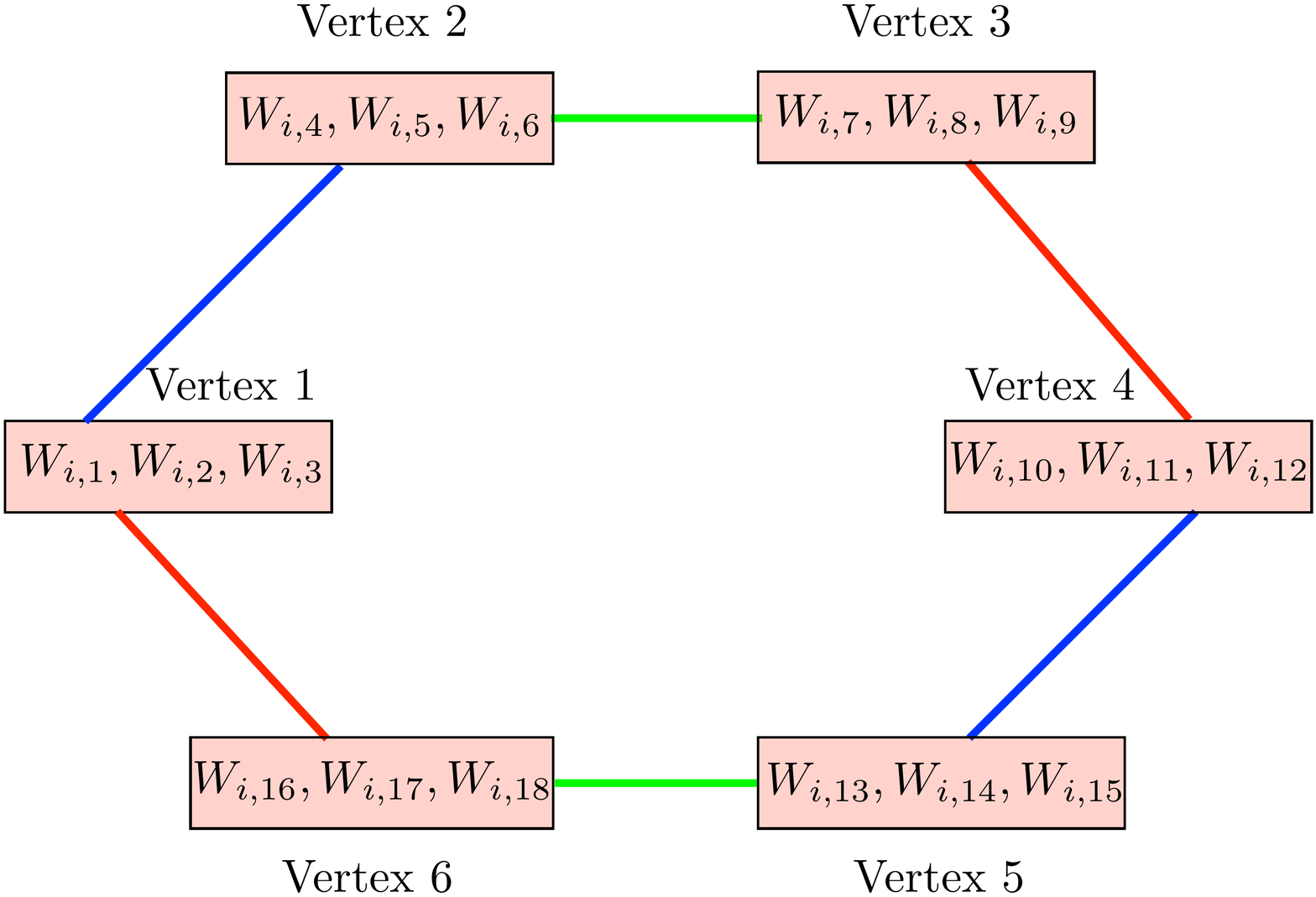}
\label{fig: RS graph}
}
\subfigure[]{
\centering \includegraphics[width=7cm]{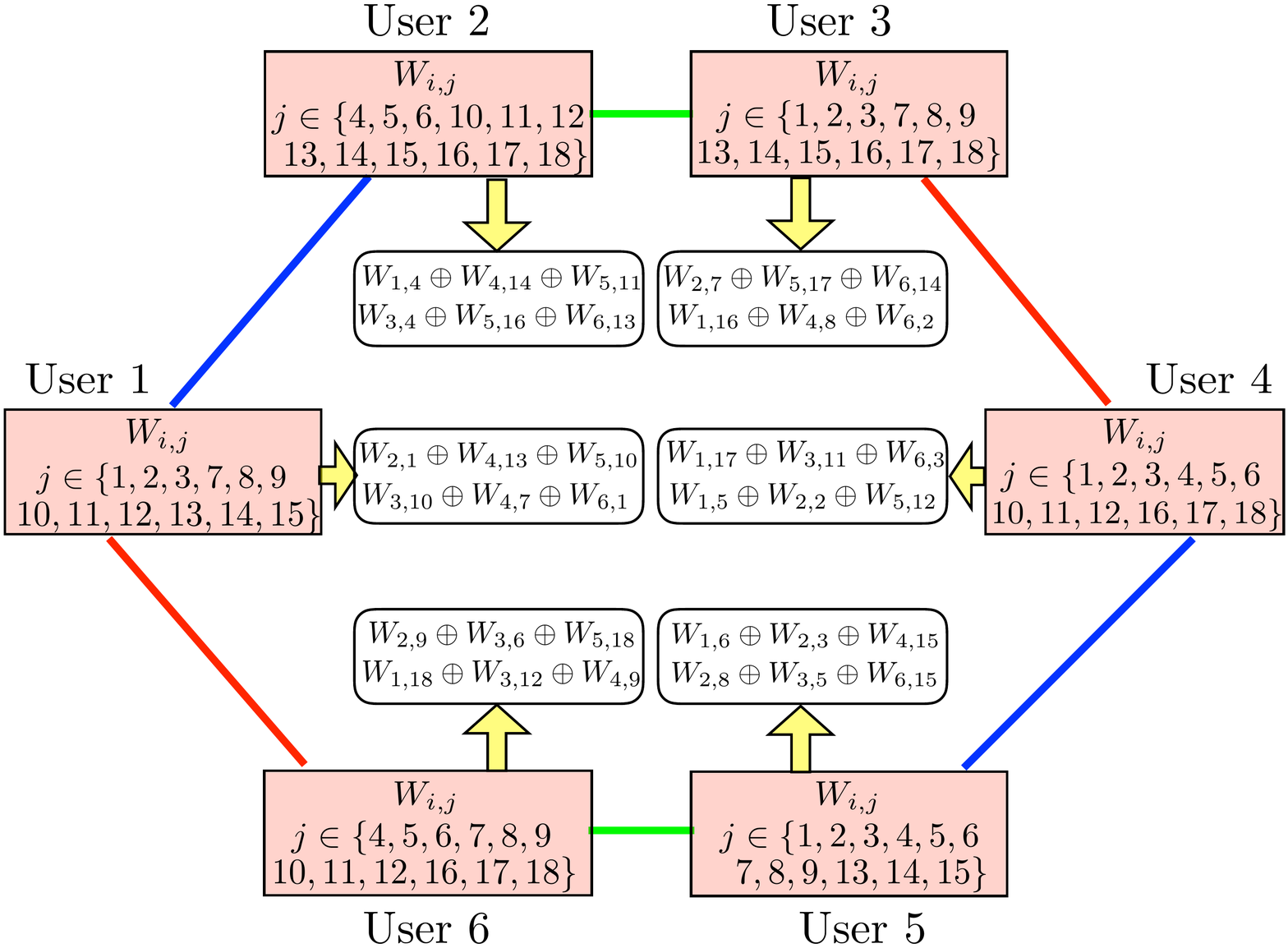}
\label{fig: D2D_RS_Cache_Example}
}
%\vspace{-0.2cm}
\caption{~\small a) A $(2,3)$-Ruzsa-Sezmeredi graph with 6 vertices which represents a coded caching scheme where $n=6$, $\frac{M}{m} = \frac{2}{3}$, and $K=n(2\gamma -1)=3n=18$. The graph can be split into $\tau=3$ induced pairwise disjoint matchings denoted by different colors in the graph, %$M_1$, $M_2$ and $M_3$,
which cover all edges of the original graph. b) An example of the proposed scheme in a %disjoint induced matching for a
D2D network with $n=m=6, K=18$ and $M=4$. Each vertex represents a user. The content of each rectangle represents cached packets.  The coded multicast packets are shown in the center of the figure. %is represented by $3$ data sets. The strong edge coloring is shown by the colors of each edge and the corresponding chromatic index $\chi_{\text{sq}}(\Gc)$ in this graph is $3$.
}
%\label{fig: D2D_RS_Cache_Example}
%\vspace{-0.4cm}
\end{figure}

An example of RS graph is shown in Fig. \ref{fig: RS graph}, which is also used in \cite{shanmugam2017coded}. We will apply Ruzsa-Szem\'eredi graph to construct a cache placement and a coded multicasting scheme in wireless D2D caching networks.
%\begin{figure}
%\centering
%\includegraphics[width=8.5cm]{RS_D2D}
%\vspace{-0.2in}
%\caption{A $(2,3)$-Ruzsa-Sezmeredi graph with 6 vertices which represents a coded caching scheme where $n=6$, $M/m = 2/3$ and $K=3n$. The graph can be split into 3 induced pairwise disjoint matchings denoted by different colors in the graph, %$M_1$, $M_2$ and $M_3$,
%which cover all edges of the original graph.} %3 multicasts of rate $1/K$, defined by each subgraph, can serve all user requests. The total rate is 1/2.}
%\vspace*{-0.2in}
%\label{fig: RS graph}
%\end{figure}
In this section, we will focus on the case %when the transmission range $r \geq \sqrt{2}$. In other words, it means
that a transmission from any node %in the network
can be received and successfully decoded by all nodes in the network (e.g., $r \geq \sqrt{2}$). In the remainder of this section, we will first introduce a motivating example and then present the general achievable scheme.

\subsection{An Example}
\label{sec: example}
Similar to \cite{shanmugam2017coded}, our proposed scheme is based on RS Graphs.
In this example,  let $m=n=6$ and $M=4$. Each file is partitioned into  %the library has $m=6$ files, denoted by $\{W_i, i \in \{1, \cdots, 6\}\}$, and there are $n=6$ users in the network. Moreover, we let $M=4$, containing
$4K$ packets with $F/K$ bits each. %in each packet.
Without loss of generality, we let user $u$ request file $u$. To find the cache placement and the coded multicasting scheme,  we build a $(2, 3)$-Ruzsa-Szem\'eredi graph $\Gc$ with $6$ vertices as shown in Fig. \ref{fig: RS graph} where each  vertex is represented by $3$ packets. %(shown in the colored packets under each node in Fig. \ref{fig: D2D_RS_Cache_Example}) as mentioned before.
In our proposed achievable scheme, we partition each file into $n(2\gamma-1)=18$ packets and denote the $j$th packet from file $i$ by $W_{i,j}, i \in \{1, \cdots 6\}, j \in \{1, \cdots, 18\}$.
The cache placement and coded multicasting scheme are shown in Fig. \ref{fig: D2D_RS_Cache_Example}, where each vertex represents a user and his cached packets are shown inside each vertex. For instance, the most left vertex, vertex 1, represents user 1 and his cached packets, which are $W_{i,j}$, where $i \in \{1, \cdots, 6\}$ and $j \in \{1,2,3,7,8,9,10,11,12,13,14,15\}$. Note that these exclude the packets represented by its neighboring vertices, vertex 2 and vertex 6, which are shown in Fig. \ref{fig: D2D_RS_Cache_Example}.

The coded multicasting scheme is shown in the center of Fig. \ref{fig: D2D_RS_Cache_Example}. For example, user $1$ transmits $W_{2,1} \oplus W_{4,13} \oplus W_{5,10}$, which are used by users $2, 4, 5$ to decode packets $W_{2,1}, W_{4,13}, W_{5,10}$, respectively. In particular, it can be seen that user $2$ has $W_{4,13}$ and $W_{5,10}$ such that he can decode $W_{2,1}$. Similarly, user $4$ and $5$ can also decode the corresponding packet in a similar manner.
In this  example, if we use the achievable scheme proposed in \cite{ji2016fundamental}, each file is split into $t{n \choose t} = 4 {6 \choose 4} = 60$ packets and the transmission rate is given by $\frac{m}{M} - 1 = \frac{6}{4} - 1 = \frac{1}{2}$. While using the proposed scheme, the transmission rate is $\frac{12}{18}=\frac{2}{3}$. Therefore,  the proposed scheme requires $K=18$  instead of $K=60$  as the scheme in \cite{ji2016fundamental}, and the achievable transmission rate of the proposed scheme is $\frac{2}{3}$  instead of  $\frac{1}{2}$ achived by the scheme in \cite{ji2016fundamental}. Hence, we  observe that while sacrificing the transmission rate  by  $(\frac{2}{3} - \frac{1}{2})/\frac{1}{2} = \frac{1}{3} = 33\%$, the proposed scheme reduces the number of packets needed per file  by $\frac{60-18}{60} = \frac{7}{10} = 70\%$.

%\begin{figure}
%\centering
%%\subfigure[]{
%\centering \includegraphics[width=8.5cm, height=6.0714cm]{Caching_Coded_Multicasting}
%%}
%%\subfigure[]{
%%\centering \includegraphics[width=7cm, height=5.1cm]{Strong_Edge_Coloring_v2}
%%\label{fig: Strong_Edge_Coloring}
%%}
%\vspace{-0.2cm}
%\caption{~\small An example of the proposed scheme in a %disjoint induced matching for a
%D2D network with $n=m=6, K=18$ and $M=4$. Each vertex represents a user. The content of each rectangle represents cached packets.  The coded multicasted packets are shown in the center of the figure. %is represented by $3$ data sets. The strong edge coloring is shown by the colors of each edge and the corresponding chromatic index $\chi_{\text{sq}}(\Gc)$ in this graph is $3$.
%}
%\label{fig: D2D_RS_Cache_Example}
%\vspace{-0.4cm}
%\end{figure}

\subsection{General Achievable Scheme}
\label{sec: RS scheme}

In this section, we generalize the deterministic caching and coded delivery
scheme illustrated in Section \ref{sec: example} to the general case of  $m$, $n$ and $M$.

\begin{itemize}
\item {\bf Building the Ruzsa-Szem\'eredi graph:} We build a  Ruzsa-Szem\'eredi graph with $n$ vertices, each with a degree of $n-t$. This graph consists of $\tau$ pairwise disjoint induced matchings and each  has $\gamma$ edges.

\item {\bf Cache Placement Phase:}
The cache placement scheme is closely related to the scheme in \cite{shanmugam2017coded} and is designed according to the Ruzsa-Szem\'eredi graph.
 Each file is divided into $K = n (2\gamma-1)$ packets,  labeled by $\{W_{i,j}, i=1, \cdots, m, j = 1, \cdots, K\}$. Each vertex in the RS graph represents $2\gamma-1$ distinct packets.
Node $u$ caches  packets corresponding to those vertices that are not adjacent to itself and  packets corresponding to the vertex itself.

\item {\bf Delivery Phase:}
As a consequence of the caching scheme described above, any subset of $2\gamma-1$ nodes
belonging to a disjoint matching of size $\gamma $ in $\Uc = \{1, \ldots, n\}$ has the property that
they share $2\gamma-1$ packets from each file.
Consider one such subset.
For any file requested by the remaining $(2\gamma)$-th node, by construction, there are $2\gamma-1$ packets shared by the other $2\gamma-1$ nodes
and needed by  the $(2\gamma)$-th node. Therefore, each node in every disjoint matching %of size $\gamma$
has $2\gamma-1$ packets that are useful for the remaining $2\gamma-1$ nodes.
Furthermore, such sets of packets are disjoint (empty pairwise intersections).
For delivery, %for all subsets of $2\gamma$ nodes,
in each disjoint matching, each node computes the XOR of its $2\gamma-1$ useful
packets and multicasts it to all other nodes in this disjoint matching. In this way, for every multicast transmission exactly $2\gamma-1$ nodes will be able to decode
a useful packet using ``interference cancellation'' based on their cached content.

\end{itemize}

The achievable rate is given in the following.

\begin{theorem}
\label{theorem: 3}
Let $m,n,M$ be the library size, number of users and the cache size per user, respectively. For $r \geq \sqrt{2}$, $n = \Lambda^z$, where $\Lambda$ is any positive integer such that $z \geq 2\Lambda$, and
\be
\label{eq: M_RS}
M = 2mn^{-\frac{1}{2\Lambda^4\ln \Lambda}},
\ee
and let $t = \frac{nM}{m} \in \mathbb{Z}^+$, the following rate is achievable:
\be
\label{eq: theorem 3}
R^{\rm RS}(M) = \frac{\tau}{n} \frac{2\gamma}{2\gamma-1}
\ee
with the requirement that $K = K^{\rm RS} = (2\gamma-1)n$.
\hfill  $\square$
\end{theorem}
\begin{IEEEproof}
Theorem \ref{theorem: 1} is proved in Appendix \ref{sec: proof of theorem 1}.
\end{IEEEproof}

When we consider the asymptotic regime as $n$ becomes large,  we obtain the following corollary.
\begin{corollary}
\label{corollary: 1}
For $r \geq \sqrt{2}$, when $n \rightarrow \infty$, $K = O\left(n^{2-\delta}\right)$ and $M = 2mn^{-c_1\delta \exp\left(-c_2/8\right)}$, where $c_1,c_2$ are some positive constants, let $t = \frac{nM}{m} \in \mathbb{Z}^+$, the following rate is achievable:
\be
\label{cor 2}
R^{\rm RS}(M) \leq n^{\delta} + o(n^\delta),
\ee
where $\delta = \frac{2\ln 10.5}{\ln \Lambda}$.
\hfill  $\square$
\end{corollary}
%\begin{IEEEproof}
%Corollary \ref{corollary: 1} is proved in Appendix \ref{sec: proof of corollary 1}.
%\end{IEEEproof}

Corollary \ref{corollary: 1} can be proved using the value of $\tau$ given in Appendix \ref{sec: proof of theorem 1} and the relation $\gamma = O\left(n/\tau\right)$.
%The proof of Corollary \ref{corollary: 1} is omitted due to space limitation.
From Corollary \ref{corollary: 1}, we can see that when $\Lambda$ is large enough, or equivalently, $n$ is large enough, $\delta$ can be arbitrarily small. In other words, it can be computed from (\ref{cor 2}) that $R^{\rm RS}(M) = (10.5)^{2\log_\Lambda n}$.
The throughput achieved by the proposed achievable scheme is given as follows.
\begin{corollary}
\label{corollary: 3}
Let $C_r$ be the constant link rate under the protocol model. For $r \geq \sqrt{2}$,  the per user throughput is given by
\be  \label{cor 3}
T^{\rm RS}(M) = \frac{C_r}{R^{\rm RS}(M)} \buildrel n\rightarrow \infty \over = C_{\!\!\sqrt{2}} \cdot n^{-\delta} + o\left(n^{-\delta}\right),
\ee
where $R^{\rm RS}(M)$ and $\delta$ are given by (\ref{eq: theorem 3}) or (\ref{cor 2}), is achievable.
\hfill  $\square$
\end{corollary}
\begin{IEEEproof}
This corollary can be proved by using the same procedure as the proof of Corollary \ref{corollary: 11}.
% Following (\ref{eq: throughput}), %Definition \ref{useful-throughput-i},
% in order to deliver $F R^{\rm RS}(M)$ coded bits without spatial reuse (at most one active link transmitting at any time),
%we need $D = FR^{\rm RS}(M)/C_r$ channel uses. Therefore, we obtain (\ref{cor 1}).
\end{IEEEproof}

\subsection{Comparison to Other Schemes}

Due to the sub-quadratic packetization with the number of users $n$, which is significantly lower than that of the hypercube based approach proposed in Section \ref{sec: Hypercube Caching Network Approach}, the achievable rate of the RS graph based design given by (\ref{eq: theorem 1}) and (\ref{cor 2}) for $r \geq \sqrt{2}$ is obviously worse than the achievable rate of the hypercube based design given in (\ref{eq: R small n}) and the original design \cite{ji2016fundamental}, %. In other words,  $R^{\rm RS}$ grows as $n$ increases
when $n$ increases and $m$ and $M$ are fixed. However, there is still a significant gain in terms of transmission rate compared to the conventional uncoded unicasting scheme in some parameter regimes.
In the following, we compare the transmission rates of the uncoded scheme, $R^{\rm u}$, and the RS Graph scheme, $R^{\rm RS}$ where $m$, $M$ and $n$ are the same for both schemes. By solving (\ref{eq: M_RS}) for $n$ we obtain
\be
\label{eq: n comp}
n = \left( 2\frac{m}{M}\right) ^ {2\Lambda^4 \ln \Lambda}
% = e^{2\Lambda^4 \ln\left( 2\frac{m}{M}\right)\ln \Lambda}
\ee
which better defines the constraint relating $n$ to $M/m$ for the this scheme based on the proposed RS graph design. For the purposes of comparison, (\ref{eq: n comp}) holds for both the uncoded and RS schemes. %since we are assume $n$, $m$ and $M$ are the same for both schemes.
The transmission rate for the uncoded scheme is
\be
R^{\rm u} = n\left( 1-\frac{M}{m} \right) = \left( 2\frac{m}{M}\right) ^ {2\Lambda^4 \ln \Lambda}\left( 1- \frac{M}{m} \right).
\ee
Furthermore, the transmission rate of the RS scheme given by (\ref{eq: theorem 1}) is
%\be
%R^{\rm RS} = 2\frac{2\gamma}{2\gamma -1} n^\delta = 2\frac{2\gamma}{2\gamma -1}\left( 2\frac{m}{M}\right) ^ {2\Lambda^4 \ln 10.5}
%\ee
\be
R^{\rm RS} = \frac{2\gamma}{2\gamma -1} n^\delta = \frac{2\gamma}{2\gamma -1}\left( 2\frac{m}{M}\right) ^ {4\Lambda^4 \ln 10.5}
\ee
% where we use the exact value of the upper bound of $R^{\rm RS}$ to avoid the order notation for the purposes of comparison.
Then we obtain %comparing the two rates it is clear that
%\be
%\label{eq: RS uncoded comp}
%\frac{R^{\rm u}}{R^{\rm RS}} = \frac{1}{2}\frac{2\gamma-1}{2\gamma}\left( 2\frac{m}{M}\right) ^ {2\Lambda^4 \left(\ln \Lambda - 2\ln10.5 \right)}\left( 1- \frac{M}{m} \right).
%\ee
\be
\label{eq: RS uncoded comp}
\frac{R^{\rm u}}{R^{\rm RS}} = \frac{2\gamma-1}{2\gamma}\left( 2\frac{m}{M}\right) ^ {2\Lambda^4 \left(\ln \Lambda - 2\ln10.5 \right)}\left( 1- \frac{M}{m} \right).
\ee
Notice that, (\ref{eq: RS uncoded comp}) is only dependent on $M$, $m$ and $\Lambda$ and not dependent on $z$. This occurs because $z$ is a function of $M$, $m$ and $\Lambda$ and for these derivations $z$ is essentially substituted with $z = 2\Lambda^4\ln \left( 2\frac{m}{M}\right)/\ln \Lambda$. This %amazing
result demonstrates that for a fixed %small
$m/M$ and $\Lambda > 10.5^2$, by using $\tau = \Lambda^{\frac{2\Lambda^4\ln \left(2\frac{m}{M}\right)}{\ln \Lambda}\left(1+\frac{2\ln 10.5}{\ln \Lambda} + o(1)\right)}$ (see Appendix \ref{sec: proof of theorem 1}), we have\footnote{Note that the RS graph based approach cannot perform worse than the uncoded unicasting by the construction. The reason we need the condition that $\Lambda > 10.5^2$ is due to the fact that the $\tau$ obtained from \cite{shanmugam2017coded,alon2012nearly} and used in (\ref{eq: theorem 1}) is a sufficient condition and may not be necessary.} %there is a significant reduction in rate compared to an uncoded D2D caching network. Furthermore, for a constant $m$ and $M$,
\be
\lim\limits_{n\to \infty} \frac{R^{\rm u}}{R^{\rm RS}}= \lim\limits_{\Lambda\to \infty} \frac{R^{\rm u}}{R^{\rm RS}} =\infty,
\ee
which shows a significant gain in terms of transmission rate of the RS graph approach compared to uncoded unicasting scheme.
\begin{remark}
It can be seen that the idea of designing the decentralized coded caching approach discussed in Section \ref{sec: decentralized hypercube} can also be extended to the RS graph coded caching methods. The major difference is that the packet sets, each of which will be randomly and uniformly cached by users, are constructed based on the RS graph. The delivery procedure is similar to that in Section \ref{sec: decentralized hypercube}.
\end{remark}
%The above comparison holds because $\gamma \geq 1$ (in fact, $\gamma = \Theta \left( n \right)$)\footnote{When, $m$ and $M$ are constant, then $z = 2\Lambda^4\ln \left( 2\frac{m}{M}\right)$ and the number of edges of the RS graph is $\Theta \left( n^2 \right)$. The number of induced matchings to cover the RS graph is $\tau = \Theta \left( n \right)$ and therefore the number of edges in each induced matching, $\gamma$, is $\Theta \left( n \right)$.} and $n>\Lambda$.

%%%%%%%%%%%%%%%%%%%%%%%%%%%%%%%%%%%%%
%%%%%%%%%%%%%%%%%%%%%%%%%%%%%%%%%%%%%
%While the RS scheme demonstrates a reduction in rate compared to the uncoded scheme, we can also show that there is cost to having a sub-quadratic packetization when comparing the transmission rates of \cite{ji2016fundamental} and the RS scheme. More specifically,
%
%\be
%\frac{R^{\rm RS}}{R'} = 2\frac{2\gamma}{2\gamma -1}\frac{M}{m-M}\left( 2\frac{m}{M}\right) ^ {2\Lambda^4 \ln 10.5}
%\ee
%and
%\be
%\lim\limits_{n\to \infty} \frac{R^{\rm RS}}{R'}= \lim\limits_{\Lambda\to \infty} \frac{R^{\rm RS}}{R'} =\infty.
%\ee
%This is expected since the RS scheme requires at most $n^2$ packets per file compared to $K'=t { n \choose t}$ packets per file in \cite{ji2016fundamental}. When considering either scheme there is a clear trade-off between the number of packets per file and the transmission rate. Also, it is obvious that as the number of users becomes large that
%
%\be
%\lim\limits_{n\to \infty} \frac{K^{\rm RS}}{K'}= \lim\limits_{\Lambda\to \infty} \frac{K^{\rm RS}}{K'} = 0.
%\ee
%%%%%%%%%%%%%%%%%%%%%%%%%%%%%%%%%%%%%
%%%%%%%%%%%%%%%%%%%%%%%%%%%%%%%%%%%%%

\section{Achievability with Spatial Reuse} %for $r < \sqrt{2}$}
\label{sec: SR}
In this section, we choose (e.g., reduce) the  transmission range $r$ in order to have localized D2D communication such that some spatial reuse is allowed. In contrast to the regime where $r \geq \sqrt{2}$, in this case, we also need to design a non-trivial transmission policy to schedule concurrent active D2D transmissions.
Similar to the scheduling schemes  in \cite{ji2016fundamental}, the proposed policy is based on {\em clustering}: the network is divided into clusters of equal size $g_c$ users. Users are  allowed to receive messages only from nodes in the same cluster.\footnote{Note that this condition can be relaxed by using the similar communication scheme base on ITLinQ \cite{Naderializadeh2014ITLinqCaching}.}
 Therefore, each cluster is treated as a smaller network compared to the entire network.
Assuming that $g_c M \geq m$,\footnote{If the condition $g_c M \geq m$ is not satisfied,
we can choose a larger transmission range such that this condition is feasible.}
the total cache size of each cluster is sufficient to store the entire file library.
Under this assumption, the cache placement and delivery schemes introduced in Sections \ref{sec: hc general scheme} and \ref{sec: RS scheme} can be applied to each cluster. Hence, it can been seen that the transmission rate in each cluster is given by either

\be
\label{eq: Rc}
R_c^{\rm hc}(M) = \frac{t_{\rm c}}{t_{\rm c}-1}\frac{m}{M}\left(1-\frac{M}{m}\right),
\ee
where we define $t_{\rm c} \eqdef g_{\rm c}M/m$,
or
\be
\label{eq: Rc 1}
R_{\rm c}^{\rm RS}(M) = \frac{\tau}{g_{\rm c}} \frac{2\gamma}{2\gamma-1}  \buildrel g_{\rm c} \rightarrow \infty \over = g_{\rm c}^{\delta} + o(g_{\rm c}^\delta),
\ee
where $t_{\rm c} = g_{\rm c}M/m$, $g_{\rm c}=\Lambda^z$ and $\tau$, $\gamma$ and $\delta$ are given in Theorem \ref{theorem: 3}. As mentioned in Remark \ref{remark 1}, a straightforward achievable transmission policy consists of grouping the set of clusters into
$\Kc$ spatial reuse sets such that the clusters of the same reuse set do not interfere and can be activated simultaneously.\footnote{Note that the interference management in this paper is based on protocol model. However, similar scheme can be designed for AWGN or fading channel model based on the condition that treating interference as noise is optimal (ITLinQ \cite{Naderializadeh2014ITLinqCaching}).} In each active cluster,
a single transmitter is active per time-slot and it is received by all  nodes in the cluster, as in classical  time-frequency reuse
schemes with reuse factor $\Kc$ that are currently used in cellular networks \cite[Ch. 17]{molisch2011wireless}.
An example of a reuse set is shown in Fig.~\ref{fig: GridNetwork}. In particular, we can pick $\Kc = \left(\left\lceil\sqrt{2}(1+\Delta)\right\rceil+1\right)^2$ \cite{xue2006scaling}.

In \cite{ji2016fundamental}, we observe that there is potentially no order gain in throughput as $n \rightarrow \infty$ by using spatial reuse. The impact of spatial reuse on throughput is completely determined by the value of link rate $C_r$ and reuse factor $\Kc$. In the newly proposed schemes, since the transmission rate in each cluster is a function of the number of users, spatial reuse may achieve a higher gain in terms of throughput. Perhaps the more interesting aspect of spatial reuse in is how it can be used to reduce packetizations in the D2D caching network. In a shared link network it is obvious that dividing the network will greatly reduce packetizations, however, throughput will significantly decrease. In D2D caching networks, we see that throughput may increase or not change significantly in some parameter regimes, and therefore we can reduce the number of packets without sacrificing throughput. Note that from the achievability point of view, reducing the transmission range $r$ is equivalent to reducing $g_{\rm c}$. %Next, we consider two ways of reducing $g_{\rm c}$ and study their impact on achievable rate, throughput, and cache size requirement.
We consider the use of the hypercube and RS graph approaches separately as we investigate clustering in D2D caching networks.

\subsection{Clustering with the Hypercube Caching Approach}

The multiplicative gap between the transmission rate of $R_{\rm c}^{\rm hc}$ and $R_{\rm g}^{\rm hc}$ is given by
\be
\label{eq: G_R hc}
G_R^{\rm hc} = \frac{R_{\rm c}^{\rm hc}}{R_{\rm g}^{\rm hc}} = \frac{g_{\rm c}}{n}\frac{t-1}{t_{\rm c}-1}.
\ee
From (\ref{eq: G_R hc}), we can see that the transmission rate of each cluster decreases as $g_{\rm c}$ decreases. Furthermore, the achievable throughput of the hypercube approach with clustering %and packetization pair
is given by the following theorem.
\begin{theorem}
\label{theorem: 4}
Let $m,n,M$ be the library size, number of users and the cache size per user, respectively. For $r \leq \sqrt{2}$, $t_{\rm c} = \frac{g_{\rm c}M}{m} \geq 2$, $c(t_{\rm c}-1) = \frac{m}{M} - 1$ and $\frac{m}{M}, t_{\rm c}, c \in \mathbb{Z}^+$, the following per user throughput $T_{\rm c}^{\rm hc}(M)$ is achievable:
\be
\label{eq: Tc K}
T_{\rm c}^{\rm hc}(M) = \frac{C_r}{\mathcal{K}}\frac{M}{m-M}\frac{t_{\rm c}-1}{t_{\rm c}}
\ee
with the requirement of $K = K_{\rm c}^{\rm hc} = \left(\frac{m}{M}\right)^{t_{\rm c}}$.
\end{theorem}
\begin{IEEEproof}
By using (\ref{eq: Tc}), we can obtain
\be
T_{\rm c}^{\rm hc}(M) = \frac{C_r}{\mathcal{K}}\frac{1}{R_{\rm c}^{\rm hc}(M)}=\frac{C_r}{\mathcal{K}}\frac{M}{m-M}\frac{t_{\rm c}-1}{t_{\rm c}},
\ee
where $R_{\rm c}^{\rm hc}(M)$ is given by (\ref{eq: Rc}).

Using (\ref{eq: K}) and replacing $t$ by $t_{\rm c}$, we get
\be
\label{eq: cluster K}
K_{\rm c}^{\rm hc} = \left(\frac{m}{M}\right)^{t_{\rm c}}
\ee
\end{IEEEproof}

The multiplicative gap of $T_{\rm g}^{\rm hc}$ and $T_{\rm c}^{\rm hc}$ throughput is given by
\be
G_T^{\rm hc} = \frac{T_{\rm c}^{\rm hc}}{T_{\rm g}^{\rm hc}} = \frac{C_r}{C_{\!\! \sqrt{2}} \; \Kc}\frac{1}{G_R^{\rm hc}} = \frac{C_r}{C_{\!\!\sqrt{2}}\;\Kc}\frac{n}{g_{\rm c}}\frac{t_{\rm c}-1}{t-1}
\ee
\begin{remark}
Note that the throughput has increased by using the clustering scheme compared to the scheme for $r \geq \sqrt{2}$ if $G_T^{\rm hc} > 1$ or in other words
\be
\frac{C_r}{C_{\!\!\sqrt{2}} \; \mathcal{K}}> \frac{g_{\rm c}}{n}\frac{t-1}{t_{\rm c}-1}.
\ee
This relationship can be further simplified if $g_{\rm c}M\gg m$ and clustering will improve the throughput if $C_r/\mathcal{K}>C_{\!\!\sqrt{2}}$. This provides a similar result to that was shown in \cite{ji2016fundamental}. The exact relationship between link rate, $C_r$, and transmission range, $r$, %is %most likely dependent
depends on the physical channel model and wireless network design. However, $C_r$ may not decrease as $r$ decreases. Hence, clustering has the potential to increase throughput. In addition, from (\ref{eq: cluster K}), clustering can also be used for the purpose of reducing the number of file partitions as also shown in \cite{ji2016fundamental}. For a cluster with $g_{\rm c}$ users, the number of file partitions is
\be
K_{\rm c}^{\rm hc} = \left(\frac{m}{M}\right)^{t_{\rm c}}=\left(\frac{m}{M}\right)^{g_{\rm c}M/m},
\ee
and the ratio of the number of file partitions with and without clustering is
\be
\frac{K_{\rm g}^{\rm hc}}{K_{\rm c}^{\rm hc}}=\left(\frac{m}{M}\right)^{\frac{M}{m}(n-g_{\rm c})},
\ee
where $K_{\rm g}^{\rm hc}$ given in (\ref{eq: K}) is the number of file partitions when $r\geq\sqrt{2}$ such that no clustering is used. %and there is no clustering.
As the cluster size $g_{\rm c}$ becomes smaller, an exponential decrease in the number of file partitions is observed.
\end{remark}

\subsection{Clustering with the RS Graph Caching Approach}

In this section we first discuss the impact of clustering on packetization and throughput with the RS graph coded caching approach. We then consider two ways of reducing $g_{\rm c}$ based on the construction in \cite{shanmugam2017coded,alon2012nearly} and study their specific impact on throughput and cache size requirement. The RS graph scheme achieves the following throughput:
\begin{theorem}
\label{theorem: 2}
Let $m,n,M$ be the library size, number of users and the cache size per user, respectively. For $r < \sqrt{2}$, $K = (2\gamma-1)g_{\rm c}$, $g_{\rm c} = \Lambda^z$, where $\Lambda$ is any positive integer such that $z \geq 2\Lambda$, and $M = 2mg_{\rm c}^{-\frac{1}{2\Lambda^4\ln \Lambda}}$, let $t_{\rm c} = \frac{g_{\rm c}M}{m} \in \mathbb{Z}^+$, the per user throughput is given by:
\be
\label{eq: theorem 2}
T_{\rm c}^{\rm RS}(M) = \frac{C_r}{\Kc} \frac{g_{\rm c}}{\tau} \frac{2\gamma-1}{2\gamma}. %\frac{m}{M}\left(1-\frac{M}{m}\right).
\ee
%where $\delta = 2\frac{\ln 10.5}{\ln \Lambda}$.
%Moreover, when $t$ is not an integer, the convex lower envelope of $R(M)$, seen as a function of $M \in [0:m]$, is achievable.
\hfill  $\square$
\end{theorem}
%\begin{IEEEproof}
%Theorem \ref{theorem: 2} is proved in Appendix \ref{sec: proof of theorem 2}.
%\end{IEEEproof}
%When we consider the asymptotic regime where $n$ becomes large,
We have the following corollary for the asymptotic regime.
\begin{corollary}
\label{corollary: 2}
For $r < \sqrt{2}$, when $g_{\rm c} \rightarrow \infty$, $K = O\left(g_c^{2-\delta}\right)$, $t_c \geq 2$, and $M = 2mg_c^{-c_1\delta \exp\left(-c_2/8\right)}$, where $c_1,c_2$ are some positive constants, let $t_{\rm c} = \frac{g_{\rm c}M}{m} \in \mathbb{Z}^+$, the throughput is given by:
\be
\label{eq: corollary 2}
T_{\rm c}^{\rm RS}(M) = \frac{C_r}{\Kc} g_{\rm c}^{-\delta} + o\left(\frac{C_r}{\Kc} g_{ \rm c}^{-\delta} \right),
\ee
where $\delta = \frac{2\ln 10.5}{\ln \Lambda}$. %is an arbitrarily small positive number.
%Moreover, when $t$ is not an integer, the convex lower envelope of $R(M)$, seen as a function of $M \in [0:m]$, is achievable.
\hfill  $\square$
\end{corollary}
Since the proofs of Theorem \ref{theorem: 2} and Corollary \ref{corollary: 2} are similar to that of Theorem \ref{theorem: 4} by using Theorem \ref{theorem: 1}  and Corollary \ref{corollary: 1}, the proofs are omitted due to space limitation. From Theorems \ref{theorem: 3} and \ref{theorem: 2}, we find the multiplicative reduction in packetization from clustering with the RS scheme is
\be
\frac{K^{\rm RS}}{K^{\rm RS}_{\rm c}} = O\left( \frac{n^2}{g_{\rm c}^2} \right).
\ee

%First, we consider
\subsubsection{Decreasing $g_{\rm c}$ by reducing $z'$ while keeping $\Lambda$ the same} Let $n = \Lambda^z$ and $g_{\rm c} = \Lambda^{z'}$, where $z' < z$.
From %Corollary \ref{corollary: 1}
(\ref{cor 2}) and (\ref{eq: Rc 1}), the multiplicative gap between the transmission rate of $R_{\rm c}^{\rm RS}(M)$ and $R^{\rm RS}(M)$ is given by
\begin{eqnarray}
\label{eq: GR-1}
G_R^{\rm RS} &=& \frac{R_{\rm c}^{\rm RS}(M)}{R^{\rm RS}(M)}
= \frac{g_{\rm c}^\delta + o(g_{\rm c}^\delta)}{n^{\delta} + o(n^{\delta})} %\notag\\
%&=& \left(\frac{n}{g_c}\right)^{\delta}  + o\left(\left(\frac{n}{g_c}\right)^{\delta}\right) \notag\\
= \Lambda^{\delta(z'-z)} + o\left(\Lambda^{\delta(z'-z)} \right) %\notag\\
%&=& \Lambda^{\frac{2\ln 10.5}{\ln \Lambda}(z-z')} + o\left(\frac{2\ln 10.5}{\ln \Lambda}^{\delta(z-z')} \right) \notag\\
= 10.5^{2(z'-z)} + o\left(10.5^{2(z'-z)}\right).
\end{eqnarray}
From (\ref{eq: GR-1}), we can observe that by clustering the D2D network, the transmission rate (traffic load) can be reduced significantly. For example, if $z-z'=2$, e.g., the network is partitioned into $\Lambda^2$ clusters,  $G_R^{\rm RS} = 10.5^{-4} < 10^{-4}$. Nevertheless, due to the spatial reuse gain, by using Corollary \ref{corollary: 11} and \ref{corollary: 2}, the multiplicative gap between $T_{\rm c}^{\rm RS}(M)$ and $T^{\rm RS}(M)$ is given by
\begin{eqnarray}
\label{eq: GT-1}
G_T^{\rm RS} &=& \frac{T_{\rm c}^{\rm RS}(M)}{T^{\rm RS}(M)} = \frac{\frac{C_r}{\Kc} g_{\rm c}^{-\delta} + o\left(\frac{C_r}{\Kc} g_{\rm c}^{-\delta} \right)}{ C_{\!\!\sqrt{2}}\; n^{-\delta} + o\left( C_{\!\!\sqrt{2}} \; n^{-\delta}\right)} \notag\\
&\buildrel (a) \over =& \frac{C_r}{C_{\!\!\sqrt{2}}} \frac{1}{\Kc} 10.5^{2(z-z')} + o\left(10.5^{2(z-z')}\right), %o\left(\frac{C_{\sqrt{2}}}{C_r} \frac{1}{\Kc} 10.5^{2(z-z')}\right),
%\notag
\end{eqnarray}
where (a) is obtained by repeating the same procedure as (\ref{eq: GR-1}). From (\ref{eq: GT-1}), we can see that if $\frac{C_r}{C_{\sqrt{2}}} \frac{1}{\Kc} $ and $\Lambda$ are positive constants, then by increasing $z-z'$, e.g., increasing the number of clusters, the multiplicative gap $G_T^{\rm RS}$ between  $T_{\rm c}^{\rm RS}(M)$ and $T^{\rm RS}(M)$ can grow to infinity, which means that the gain by using the spatial reuse in D2D network can be infinite. In practice, even for finite $z-z'$, it is possible to achieve gains by using spatial reuse. For example, if $z-z' = 2$, then from (\ref{eq: GT-1}), we can see that %to achieve gains by using spatial reuse, the following condition needs to be satisfied.
\be
G_T^{\rm RS} = \frac{C_r}{C_{\sqrt{2}}} \frac{1}{\Kc} 10.5^{4},
\ee
which can be much larger than $1$. This surprising result demonstrates that spatial reuse provides significant gain in throughput even when the reuse factor $\Kc$ is relatively large and there is little increase in the link capacity from decreasing the transmission range.
%Population clustering can be performed for multiple iterations and each iteration provides a throughput gain of $\sim 10.5^4$, until the D2D loss coefficients become significant as the size of the groups decreases.

As clustering is performed, the required cache capacity $M$ increases. For this specific clustering scheme, from Theorem \ref{theorem: 1} and Theorem \ref{theorem: 2}, the following condition for cache capacity $M(g_{\rm c})$ must hold
\begin{eqnarray}
\label{eq: M}
M(g_{\rm c}) &=& 2mg_{\rm c}^{-\frac{1}{2\Lambda^4\ln \Lambda}} = 2mn^{-\frac{1}{2\Lambda^4\ln \Lambda}} \frac{g_{\rm c}^{-\frac{1}{2\Lambda^4\ln \Lambda}}}{n^{-\frac{1}{2\Lambda^4\ln \Lambda}}} \notag\\
%&=& M(n) \left(\frac{n}{g_c}\right)^{\frac{1}{2\Lambda^4\ln \Lambda}} %\notag\\
%= M(n) \Lambda^{\frac{z-z'}{2\Lambda^4\ln \Lambda}} \notag\\
&=& M(n) \exp\left(\frac{z-z'}{2\Lambda^4}\right).
\end{eqnarray}
%\be
%M\geq 2m\Big(\frac{n}{C^2}\Big)^{-\frac{1}{2\Lambda^4\ln \Lambda}}=2mn^{-\frac{1}{2C^4\ln C}}e^{\frac{1}{C^4}}.
%\ee
Note that by using (\ref{eq: M}), to achieve the promised throughput, it needs  %cache capacity
$M(g_{\rm c}) \geq M(n) \exp\left(\frac{z-z'}{2\Lambda^4}\right)$. For example, if $z-z'=2$, it requires $M(n/\Lambda^2) = M(n)e^{\frac{1}{\Lambda^4}}$.
This demonstrates that if the network is divided into $\Lambda^2$ clusters, the lower bound of $M$ increases by a factor of $e^{\frac{1}{\Lambda^4}}$, which is insignificant for  reasonablely large values of $\Lambda$.

%Next, we
\subsubsection{Decrease $g_{\rm c}$ by reducing $\Lambda$ while keeping $z$ unchanged} Let $n = \Lambda^z$ and $g_{\rm c} = (\Lambda')^{z}$, where $\Lambda' < \Lambda$.
From Corollary \ref{corollary: 1} and (\ref{eq: Rc}), the multiplicative gap between the transmission rate of $R_{\rm c}^{\rm RS}(M)$ and $R^{\rm RS}(M)$ is given by
\begin{eqnarray}
\label{eq: GR}
G^{\rm RS}_R = \frac{R_{\rm c}^{\rm RS}(M)}{R^{\rm RS}(M)}
= \frac{g_c^{\delta'} + o(g_c^{\delta'})}{n^{\delta} + o(n^{\delta})} %\notag\\
%&=& \left(\frac{n}{g_c}\right)^{\delta}  + o\left(\left(\frac{n}{g_c}\right)^{\delta}\right) \notag\\
%&=& \left(\frac{\Lambda^\delta}{(\Lambda')^{\delta'}}\right)^{z} + o\left(\left(\frac{\Lambda^\delta}{(\Lambda')^{\delta'}}\right)^{z} \right) \notag\\
= \frac{(\Lambda')^{\frac{2\ln 10.5}{\ln \Lambda'}z}}{\Lambda^{\frac{2\ln 10.5}{\ln \Lambda}z}} + o\left(\frac{(\Lambda')^{\frac{2\ln 10.5}{\ln \Lambda'}z}}{\Lambda^{\frac{2\ln 10.5}{\ln \Lambda}z}} \right) %\notag\\
= 1 + o\left(1\right).
\end{eqnarray}
From (\ref{eq: GR}), interestingly, we can see that in this case by clustering the D2D network, the transmission rate  is almost unchanged. It follows from Corollary \ref{corollary: 11} and \ref{corollary: 2} that the multiplicative gap between $T_{\rm c}^{\rm RS}(M)$ and $T^{\rm RS}(M)$ is given by
\begin{eqnarray}
\label{eq: GT}
G_T^{\rm RS} &=& \frac{T_{\rm c}^{\rm RS}(M)}{T^{\rm RS}(M)} = \frac{\frac{C_r}{\Kc} g_{\rm c}^{-\delta'} + o\left(\frac{C_r}{\Kc} g_{\rm c}^{-\delta'} \right)}{ C_{\!\!\sqrt{2}} \; n^{-\delta} + o\left( C_{\!\!\sqrt{2}} \; n^{-\delta}\right)} \notag\\
& =& \frac{C_r}{C_{\!\!\sqrt{2}}} \frac{1}{\Kc} + o\left(\frac{C_r}{C_{\!\!\sqrt{2}}} \frac{1}{\Kc} \right).
\end{eqnarray}
From (\ref{eq: GT}), we can see that similar to \cite{ji2016fundamental}, there is no fundamental cumulative gain by using both
spatial reuse and coded multicasting. Under our assumptions, spatial reuse may or may not be advantageous,
depending on whether $\frac{C_r}{\Kc}$ is larger or smaller than $C_{\!\!\sqrt{2}}$. In addition, this also depends on  how the link rate varies as a function of  communication range.
This aspect is not captured by the protocol model,  and the answer may depend on the operating frequency and appropriate channel
model of the underlying wireless network physical layer \cite{Ji2016WirelessD2D}.
%if $\frac{C_{\sqrt{2}}}{C_r} \frac{1}{\Kc} $ and $C$ are positive constant, then by increasing $z-z'$, e.g., increasing the number of clusters, the multiplicative gap $G_T$ between the $T(M)$ and $T_c(M)$ can grow to infinity, which means that the gain by using the spatial reuse in D2D network can be infinite. In practice, even for finite $z-z'$, it is possible to achieve gains by using spatial reuse.

\section{Conclusion}
\label{sec: Conclusion}

This work aims to find D2D coded caching network designs which require less packetization as compared to the state-of-the-art coded D2D caching schemes in \cite{ji2016fundamental}. We propose two new network design approaches, which are hypercube approach and Ruzsa-Szem\'eredi graph approach. In particular, the hypercube approach has been designed specifically for D2D and not based off a shared link scheme as that in \cite{ji2016fundamental}. The hypercube approach requires exponentially less packetization as compared to coded caching scheme in \cite{ji2016fundamental} and yields nearly the same per user throughput for a large number of users. The RS Graph approach was expanded from a previously studied shared link network design \cite{shanmugam2017coded} and modified for the design of D2D networks.  It requires only subquadratic packetization in terms of the number of users and achieves a near constant per user throughput. {The dramatic decrease in packetization resulted from the RS Graph approach comes at the cost of a significant decrease in throughput compared to \cite{ji2016fundamental}}. %Future work may yield other RS graph constructions with more general packetization requirements and a trad-off between packetization and transmission rate.
In addition, we demonstrate how the hypercube scheme can be expanded to design decentralized networks and discovered a clear trade-off between transmission rate and packetization.
Furthermore, we explore our new schemes by breaking up larger networks into {smaller} networks to utilize spatial reuse and increase per user throughput and also decrease the packetization, a property that is unique to D2D caching networks. %Lastly, we demonstrate how the hypercube scheme can be expanded to design decentralized networks and discovered a clear trade-off between transmission rate and packetization.
%This work represents a major step forward towards practical file packetization in D2D coded caching networks. Future work may yield other RS graph constructions with more general packetization requirements and a trade-off between packetization and transmission rate.

\appendices

\section{Proof of the Correctness of the General Hypercube Scheme}
\label{sec: Correctness}

In this section, we demonstrate the correctness of the proposed scheme by verifying that all users: 1) locally cache at most $MF$ bits; 2) only transmit content that is locally cached; 3) decode requested packets from received coded multicasts; and 4) recover the entire requested file.

For 1), each user caches a set of $(\frac{m}{M})^{t-1}$ packets of size $\frac{F}{(m/M)^t}$ bits for all $m$ files in the library. The total number of bits cached locally at each user is
\be
\left(\frac{m}{M}\right)^{t-1} \cdot \frac{F}{\left(\frac{m}{M}\right)^{t}} \cdot m=MF
\ee
which satisfies the local memory constraint for each user.

%Each user $u_i \in \mathcal{S}$ is guaranteed to have the packets cached that user $u_i$ is to transmit. This is true because for

For 2), for each packet, $W_{f_{u_s},(a_0,\ldots ,a_{t-1})}$, included in a multicast from user $u_i$, notice that $a_i = u_i \bmod \frac{m}{M}$, which follows from (\ref{eq: dlv index 1}) and (\ref{eq: dlv index 2}), and $\lfloor u_i/\frac{m}{M} \rfloor = i$ as defined for a user in a set $\mathcal{S}$ in the delivery phase. This precisely matches the definition for a packet that user $u_i$ cached during the placement phase, and therefore each user $u_i \in \mathcal{S}$ is guaranteed to have the packets cached that it is about to transmit.

For 3), first notice that, for any user set, $\mathcal{S}$, as defined in the delivery phase, a user $u_j\in \mathcal{S}$ will receive a requested packet from all users in $\mathcal{S}\setminus u_j$. To prove this consider any user $u_i\in \mathcal{S}\setminus u_j$. Notice that $(j-i)\bmod \frac{m}{M}\neq 0$ because $i\neq j$ and $i,j\in \{0,\ldots ,\frac{m}{M}-1\}$. Furthermore, by (\ref{eq: dlv index 1}) and (\ref{eq: dlv index 2}), the multicast from user $u_i$ will include a packet $W_{f_{u_j},(a_0,\ldots ,a_{t-1})}$ such that
\be
a_j=(u_j+j-i) \bmod \frac{m}{M}\neq u_j\bmod \frac{m}{M}.
\ee
This precisely defines a packet which is part of user $u_j$'s request and is not included in user $u_j$'s cache. Next, consider every other packet, $W_{f_{u_k},(a_0,\ldots ,a_{t-1})}$, included in a multicast from the user $u_i$ and notice that $a_j = u_j \bmod\frac{m}{M}$ which follows from (\ref{eq: dlv index 1}) and (\ref{eq: dlv index 2}) where $k\neq i,j$. This means user $u_j$ has these packets cached and is able to decode the uncached, requested packet from the received multicast from user $u_i$.

For 4), first consider the following useful property
\be
  \{(a+b) \bmod c : b\in\{-x,\ldots ,c-1-x\}\setminus 0\} =\{0,\ldots ,c-1\}\setminus a \bmod c
\ee
where $x\in \mathbb{Z}$ and $0\leq x \leq c-1$. This property holds because $\{b\bmod c \} = \{1,\ldots ,c-1\}$. Now consider a user group, $\mathcal{S}$, as defined by the delivery phase. By (\ref{eq: dlv index 1}) and (\ref{eq: dlv index 2}), user $u_j\in\mathcal{S}$ will receive every packet, $W_{f_{u_j},(a_0,\ldots ,a_{t-1})}$, such that $a_q$ equals some constant if $q\neq j$ and
\be
a_j\in \{(u_j+j-i)\bmod \frac{m}{M}:i\in\{0,...,t-1\}\setminus j\}.
\ee
Furthermore, recognizing that $i-j\in \{-j,...,t-1-j \}\setminus 0$ and $t=m/M$ demonstrates that
\be
a_j\in\{0,\ldots ,\frac{m}{M}-1\}\setminus a_j \bmod \frac{m}{M}.
\ee

In other words, user $u_j$ collectively receives a ``line" of packets from users $\mathcal{S}\setminus u_j$ for which $t-1$ dimensions of the ``hypercube" are held constant and one dimension varies the length of the ``hypercube". The only exception is the packet which user $u_j$ already has cached. Moreover, the constants, $\{a_0,\ldots ,a_{t-1}\}\setminus a_j$, are defined as $a_q=u_q \bmod \frac{m}{M}$. Therefore, for user $u_j$ to receive any packet, $W_{f_{u_j},(a_0,\ldots ,a_{t-1})}$, not already cached, user $u_j$ can form a multicast group with the set of users
\be
\mathcal{S}'= \{ a_q + q(\frac{m}{M}):q\in\{ 0,\ldots ,t-1 \}\setminus j \}
\ee
Given that, $a_q \in \{0,\ldots ,\frac{m}{M}-1\}$ for all $q\in \{0,...,t-1\}$, recognize that $\mathcal{S}' \cup u_j \subset \mathcal{U}$. Also, $|\mathcal{S}' \cup u_j|=t$ and there exists a user $u_k \in \mathcal{S}' \cup u_j$ such that $\lfloor u_k/\frac{m}{M}\rfloor = k$ for all $k\in \{0,\ldots,\frac{m}{M}-1\}$. Therefore, $\mathcal{S}' \cup u_j$ meets the requirements for a group, $\mathcal{S}$, as defined by the delivery phase. This proves that by forming all user sets as defined in the delivery phase, all users will recover all packets of the requested file that were not already cached.

%\vspace{-0.5cm}
\section{Proof of Theorem \ref{theorem: 3}}
\label{sec: proof of theorem 1}

To prove Theorem \ref{theorem: 3}, we first introduce the following construction of RS Graph from \cite{shanmugam2017coded,alon2012nearly}.
\begin{defn}
\label{def: RS Graph Construction}
{\bf (Graph Construction)} A graph, $\Gc(\Vc,\Ec)$, is defined such that $\Vc = [\Lambda]^z$ where $\Lambda \in \mathbb{Z}^+$, $z$ is even, and $z \geq 2\Lambda$. Let $\mu =\mathbb{E}_{x,y}[||x-y||^2_2]$, where $x$ and $y$ are sampled uniformly from $\Vc$. For a pair of vertices, $u, v\in \Vc$, $(u,v)\in \Vc$ if and only if $|\ ||u-v||^2_2-\mu|\leq z$.
\hfill $\lozenge$
\end{defn}

Given this construction, the graph consists of $\tau=n^f$ edge disjoint induced matchings and misses at most $n^k$ edges such that
%\be
%\label{eq: f}
$f=1+2\frac{\ln 10.5}{\ln \Lambda}+o(1)$,
%\ee
and
%\be
%\label{eq: k}
$k = 2-\frac{1}{2\Lambda^4\ln \Lambda}+o(1)$.
%\ee

Furthermore, given a vertex, $x \in \Vc$, the degree of $x$, defined as $d$, is bounded by \cite{shanmugam2017coded}
\be
d  = |y\in \Vc : (x,y)\in \Ec |\geq n \Big( 1-2n^{-\frac{1}{2\Lambda^4\ln \Lambda}}\Big) .
\ee

The degree of any vertex defines the number of packets that a user does not cache. Therefore, if the following holds,
$$\frac{M}{m}\geq 1-\dfrac{d_{\text{lb}}}{n}=2n^{-\frac{1}{2\Lambda^4\ln \Lambda}},$$
where $d_{\text{lb}}$ is the lower bound of $d$, then the graph construction can be used as a cache placement scheme. Hence, we can pick $M = 2mn^{-\frac{1}{2\Lambda^4\ln \Lambda}}$. Since $d \geq d_{\text{lb}} = n\left(1-\frac{M}{m}\right) = n - t$,  we have $t \in \mathbb{Z}^+$.

By the construction of the proposed scheme described in Section \ref{sec: RS scheme}, each file is partitioned into $n(2\gamma-1)$ packets, and each contains $F/K$ bits. Moreover,  for each disjoint matching in the constructed RS graph, there are $2\gamma$ transmissions, and each  has size $\frac{F}{n(2\gamma-1)}$ bits. Since the number of disjoint matching is $\tau$, there are $2\gamma \cdot \tau$ transmissions.
\be
R(M) = 2\gamma \cdot \tau \cdot \frac{F}{n(2\gamma-1)} \cdot \frac{1}{F} = \frac{\tau}{n} \frac{2\gamma}{2\gamma-1}.
\ee
Hence, we finish the proof.

%\vspace{-0.5cm}
\bibliographystyle{IEEEbib}
\bibliography{references_d2d}

\end{document}